\documentclass[aps,prb,twocolumn,showpacs,floats,fleqn,epsfig]{revtex4}
\usepackage{epsfig,psfrag,subfigure,amsopn}
\usepackage{graphicx,psfrag,subfigure}
\usepackage{amsmath,amssymb,amsbsy}
\usepackage[dvips]{color}

\newcommand\beq{\begin{equation}}
\newcommand\eeq{\end{equation}}
\newcommand\bea{\begin{eqnarray}}
\newcommand\eea{\end{eqnarray}}
\newcommand\non{\nonumber}
\newcommand\noi{\noindent}

\newcommand\ig{\includegraphics}
\newcommand\bib{\bibitem}

\newcommand\al{\alpha}
\newcommand\be{\beta}
\newcommand\ga{\gamma}
\newcommand\de{\delta}
\newcommand\De{\Delta}
\newcommand\ep{\epsilon}
\newcommand\si{\sigma}

\newcommand\lam{\lambda}
\newcommand\la{\langle}
\newcommand\ra{\rangle}

\newcommand\ta{\theta}
\newcommand\dg{\dagger}

\newcommand\ua{\uparrow}
\newcommand\da{\downarrow}
\newcommand\vk{\vec k}

\begin{document}

\title{Edge states, spin transport and impurity-induced local density
of states in spin-orbit coupled graphene}

\author{Ranjani Seshadri$^1$, K. Sengupta$^2$, and Diptiman Sen$^1$}
\affiliation{\small{
$^1$Centre for High Energy Physics, Indian Institute of Science, Bengaluru
560 012, India \\
$^2$Theoretical Physics Department, Indian Association for the Cultivation
of Science, Jadavpur, Kolkata 700 032, India}}

\date{\today}

\begin{abstract}

We study graphene which has both spin-orbit coupling (SOC),
taken to be of the Kane-Mele form, and a Zeeman field induced due to
proximity to a ferromagnetic material. We show that a zigzag
interface of graphene having SOC with its pristine counterpart hosts
robust chiral edge modes in spite of the gapless nature of the
pristine graphene; such modes do not occur for armchair interfaces.
Next we study the change in the local density of states (LDOS) due
to the presence of an impurity in graphene with SOC and Zeeman
field, and demonstrate that the Fourier transform of the LDOS close
to the Dirac points can act as a measure of the strength of the
spin-orbit coupling; in addition, for a specific distribution of
impurity atoms, the LDOS is controlled by a destructive interference
effect of graphene electrons which is a direct consequence of their
Dirac nature. Finally, we study transport across junctions which
separates spin-orbit coupled graphene with Kane-Mele and Rashba
terms from pristine graphene both in the presence and absence of a
Zeeman field. We demonstrate that such junctions are generally spin
active, namely, they can rotate the spin so that an incident
electron which is spin polarized along some direction has a finite
probability of being transmitted with the opposite spin. This leads
to a finite, electrically controllable, spin current in such
graphene junctions. We discuss possible experiments which can probe
our theoretical predictions.
\end{abstract}

\maketitle

\section{Introduction}
\label{sec:intro}

The last several years have witnessed a tremendous amount of
research on graphene, both theoretical and
experimental~\cite{been08,neto09,rev3,rev4,rev5}. Graphene is a
two-dimensional hexagonal lattice of carbon atoms in which the $\pi$
electrons hop between nearest neighbors. At half-filling, the
spectrum is gapless at two points (called $\vec K$ and $\vec K'$) in
the Brillouin zone, and the energy-momentum dispersion around both
those points has the Dirac form $E_{\vk} = \hbar v |\vk|$, where $v
\simeq 10^6 m/s$ is the Fermi velocity. The Dirac nature of the electrons
gives rise to many interesting properties of this material, such as Klein
tunneling through a barrier~\cite{katsnelson06}, novel effects of crossed
electric and magnetic fields~\cite{lukose07}, qualitatively different
transport characteristics of superconducting graphene
junctions~\cite{beenakker1,bhattacharjee06,beenakker2,maiti07}, possibility of
multichannel Kondo physics \cite{ksen3,kond1,kond2,kond3,kond4}, interesting
power laws in the local density of states (LDOS) induced by an
impurity~\cite{cheianov06,mariani07,bena08,bena09}, and atomic collapse
in the presence of charged impurities~\cite{levitov07}.

Recent years have also seen extensive research on topological
systems~\cite{hasan,qi}. These systems have a bulk spectrum which is
gapped; however, the topological properties of the bulk states
ensure, via bulk-boundary correspondence, that the boundary
(namely, the edge for a two-dimensional system like graphene) has
gapless states. The number of species of gapless states is given by
a topological invariant which can be calculated from the bulk
spectrum. While pristine graphene is gapless in the bulk and is
therefore not topological, it can be made to undergo a transition to
a topological phase with a non-zero Chern number by adding an
appropriate $s^z$ conserving spin-orbit coupling (SOC)
\cite{kane05}. Experimentally a SOC may be induced in graphene in
various ways, such as by placing it in proximity to a
three-dimensional topological insulator such as $\rm Bi_2
Se_3$~\cite{kou13,zhang14} or by functionalizing it with
methyl~\cite{zollner15}. Two models of SOC have been discussed in
the literature: Kane-Mele~\cite{kane05} and Rashba~\cite{dutreix14}.
The Kane-Mele type opens a gap and makes the system topological
while the Rashba type does not open a gap and therefore does not
make it topological; consequently, in this work, we shall deal
mostly with the former type of SOC. In addition, it is also
interesting to consider the effects of an effective magnetic field
with a Zeeman coupling to the spin of the electron. Such a coupling
can arise if a ferromagnetic material is placed in proximity to the
graphene~\cite{haugen08,yang13,song15}; the magnetization of the
ferromagnetic material will have only a Zeeman coupling to the
electron spin (no orbital coupling) provided that the direction of
the magnetization lies in the plane of the graphene. To the best of
our knowledge, edge states, impurity effects, and spin transport in
systems constituting spin-orbit and Zeeman coupled graphene have not
been studied in detail earlier.

In this work, we shall study the nature of edge states, the effects of magnetic
and non-magnetic impurities, and spin transport in junctions involving
spin-orbit coupled graphene both in the presence and absence of a Zeeman
coupling term. The pristine graphene, in our work, will be modeled by a
tight-binding lattice Hamiltonian with nearest neighbor hopping on a
hexagonal lattice~\cite{neto09}
\beq H ~=~ -~\ga ~\sum_{{\vec i}, {\vec j}} ~\sum_{\al = \ua,\da}~
(c^\dg_{{\vec i},\al} ~c_{{\vec j},\al} ~+~ H. c.), \label{ham1} \eeq
where the sum over ${\vec i}, ~{\vec j}$ goes over the nearest neighbors, 
the hopping amplitude $\ga \simeq 2.8 ~eV$, the nearest neighbor spacing
is $d \simeq 0.14 ~nm$, and $\al$ denotes the spin component in,
say, the $z$ direction. (We will set $\hbar =1$ and $\ga = 1$ unless
mentioned otherwise). The hexagonal lattice has unit cells which
consist of two sites; we denote the upper and lower sites, belonging
to sublattices $A$ and $B$, as $a_{\vec n}$ and $b_{\vec n}$
respectively. We introduce the Pauli matrices $\vec \si$ with $\si^z
= \pm 1$ denoting sites on the $A$ and $B$ sublattices respectively.
The midpoint of a unit cell labeled as $\vec n$ is located at ${\vec
n} ~=~ \sqrt{3} d ~(n_1 + \frac{1}{2} n_2, ~\frac{\sqrt{3}}{2}
n_2)$, where $n_1, ~n_2$ take integer values. The spanning vectors
of the lattice are ${\vec M_1} = \sqrt{3} d (1/2, \sqrt{3}/2)$ and
${\vec M_2} = \sqrt{3} d (1/2, - \sqrt{3}/2)$. The reciprocal
lattice vectors can be chosen to be ${\vec G_1} = (4\pi/3d)
(\sqrt{3}/2, 1/2)$ and ${\vec G_2} = (4\pi/3d) (\sqrt{3}/2, -1/2)$.
As is well-known, such a model leads to an energy dispersion $\pm
E_{\vec k}$ where
\bea E_{\vk} &=& \ga |1 + e^{i \vk \cdot {\vec M_1}} + e^{-i \vk \cdot
{\vec M_2}}| \label{ek} \\
&=& \ga [ 3 + 2 \cos (\sqrt{3} k_xd) + 4 \cos (\frac{\sqrt{3}
k_xd}{2}) \cos (\frac{3 k_y d}{2}) ]^{1/2}. \non \eea
The two bands touch each other at two inequivalent points; these are
the well-known $\vec K$ and $\vec K'$ with wave vectors $(\pm
4\pi/(3\sqrt{3}d), 0)$. Around these points, the effective
low-energy continuum theory of graphene electrons takes the form of
a $(2+1)$-dimensional Dirac Hamiltonian with \beq H_1 ~=~\sum_{\vec
k} \psi_{\vec k}^\dg [ v ~( \tau^z \si^x k_x ~-~ \si^y k_y)]
\psi_{\vec k}, \label{dirham1} \eeq where $v = 3\ga d/2$ is the
Fermi velocity, $\tau^z = \pm 1$ at ${\vec K} ~({\vec K'})$
respectively (these are called valleys), and $\psi_{\vec k} \equiv
\psi_{\vec k}^{\si \tau s}$ denote eight-component electron
annihilation operators with the components corresponding to
sublattice ($\si$), valley ($\tau$), and spin ($s$) degrees of
freedom. Equation \eqref{dirham1} is the Dirac Hamiltonian and the
dispersion is given by $E_{\vk}^\pm = \pm v | \vk|$, with a
four-fold degeneracy due to the valley and spin degrees of freedom.

\begin{figure}[htb] \ig[width=3.4in]{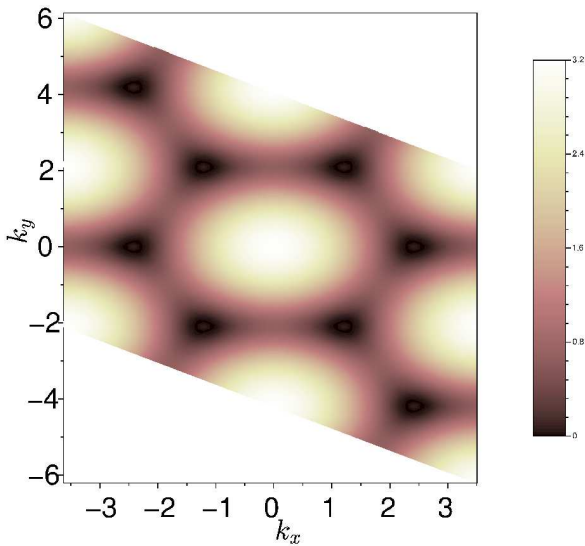}
\caption[]{Contour plot of energy-momentum dispersion for 
graphene with a SOC of strength $t_2 = 0.05$ and a Zeeman field $b_x = 0.2$. 
$E, ~t_2$ and $b_x$ are in units of $\ga$, while $k_x$ and $k_y$ are in units 
of $1/d$.} \label{fig02} \end{figure}

The presence of the SOC, taken to be of the
Kane-Mele type, and the Zeeman term arising out of proximity to a
magnetic strip will be modeled at a lattice level by \bea H_{so} &=&
i t_2 ~\sum_{{\vec i}, {\vec j}} ~\nu_{{\vec i},{\vec j}}~ (c_{{\vec
i},\ua}^\dg c_{{\vec j},\ua} ~-~ c_{{\vec i},\da}^\dg
c_{{\vec j},\da}), \label{so} \\
H_Z &=& - \sum_{\vec i}~ b_j ~c_{{\vec i},\al}^\dg s_{\al \be}^j
c_{{\vec i},\be}, \label{z} \eea
where $t_2$ denotes the strength of the SOC,
the sum over ${\vec i}, ~{\vec j}$ goes over next-nearest neighbors,
$\nu_{{\vec i},{\vec j}} = 1 ~(-1)$ if the electron makes a left
(right) turn to go from site ${\vec j}$ to ${\vec i}$ through their
common nearest neighbor, and we have taken the vector ${\vec b} =
(b_x,b_y,b_z)$, which measures the strength of the effective Zeeman
field, to include factors like the coupling to the magnetization of
a proximate ferromagnetic material and the Bohr magneton. It is easy
to see that Eqs.~\eqref{so} and \eqref{z} along with Eq.~\eqref{ham1},
lead to the continuum Hamiltonian near the Dirac points
\beq H_2 = \sum_{\vec k} \psi^\dg_{\vec k} [v ( \tau^z \si^x k_x -
\si^y k_y) + \De_{so} \tau^z \si^z s^z - {\vec b} \cdot {\vec s}]
\psi_{\vec k}, \label{dirham3} \eeq where $\Delta_{so} =
3\sqrt{3}t_2$. The energy-momentum dispersion following from
Eqs.~\eqref{ham1}, \eqref{so}, and \eqref{z} is shown in
Fig.~\ref{fig02}. In what follows, we shall use Eqs.~\eqref{ham1},
\eqref{so}, and \eqref{z} for all numerical and analytical
computations done at the lattice level and use Eq.~\eqref{dirham3}
for analyzing the continuum Dirac theory for the system.

The main results that we obtain from such an analysis are the
following. First, we study the edge states between pristine graphene
(Eq.~\eqref{ham1}) and graphene with SOC (Eq.~\eqref{so}) and
demonstrate the existence of robust chiral edge modes provided that
they are separated by a zigzag edge. No such modes exist for an
armchair edge. This result is in sharp contrast to the edge modes
between graphene with SOC and vacuum studied
earlier~\cite{kane05,lado15} where such modes exist both for
armchair and zigzag edges. We also show via an exact analytical
solution that the robustness of these edge states, in spite of the
presence of the gapless pristine graphene, is due to the fact that
the characteristic decay length of these modes vanishes in the limit
$t_2$ (or $\Delta_{so}$) $\to 0$; this behavior is in contrast to
the usual divergence of the decay length edge modes with vanishing
gap in the bulk. Second, we study spin-orbit coupled graphene in the
presence of both single and distributed impurity (impurities) in the
weak coupling limit using a $T$-matrix formalism. We compute the
energy resolved LDOS and use it to show that the width of the peaks
in the Fourier transform of the LDOS provide a direct signature of
the magnitude of the SOC. We also study a specific set of
distributed impurities and show that the corresponding LDOS reveals
a destructive interference effect which provides a direct signature
of the Dirac nature of graphene electrons. Finally, we study the
effect of magnetic impurities on the LDOS and show that they result
in a much weaker change in LDOS as compared to charged impurities.
Third, we study junctions of graphene with SOC in the form of both
Kane-Mele (Eq.~\eqref{so}) and a Rashba term given by $H_R=
\sum_{\vec k} \psi_{\vec k}^\dg [ \lambda_R (\tau^z s^x \si^y-s^y
\si^x)]\psi_{\vec k}$, both in the presence and absence of $H_{Z}$,
with pristine graphene. We show that such junctions are necessarily
spin active in the sense that electrons of a definite spin
approaching a junction may reflect from it with a different
direction of the spin. We also demonstrate that this property of
graphene junctions may be used to generate finite, electrically
controllable, spin currents and thus can provide a starting step
towards applications of such junctions in spintronics. We note that,
to the best of our knowledge, the presence of robust edge states,
the use of LDOS in the presence of impurities to estimate the
strength of the SOC, and the spin active nature of graphene
junctions leading to finite, electrically controllable, spin
currents in spin-orbit coupled graphene junctions have not been
discussed in the literature. We also note that some aspects of
Kane-Mele SOC, edge states and spin transport have been studied
recently in buckled honeycomb systems such as silicene, germanene
and stanene \cite{rachel14,liu11}; we expect our analysis
demonstrating spin active junctions and leading to electrically
controllable spin currents to hold for these materials as well (with
minor modifications to take into account the gapped Dirac spectrum of
these materials).

The plan of the paper is as follows. In Sec.~\ref{sec:edgesec}, we
discuss the physics of the edge states in graphene. This is followed
by a discussion of the LDOS due to the presence of an impurity
(impurities) in spin-orbit coupled graphene in Sec.~\ref{sec:imp}.
Next, we discuss the spin active nature of graphene junctions in
Sec.~\ref{sec:junc} and compute the spin current in several possible
junction geometries. Finally, we discuss possible experiments,
summarize our main results, and conclude in Sec.~\ref{sec:diss}.

\section{Edge states at junction of two regions}
\label{sec:edgesec}

In this section, we study localized states at the edge between
pristine graphene and graphene with SOC described by Eq.~\eqref{so}
and demonstrate the presence of exponentially localized chiral edge
states which propagate as plane wave along the edge. We will first
consider a zigzag edge.

It is known that a zigzag edge which lies between pristine graphene
and vacuum hosts edge states for a finite range of the momentum
along the edge~\cite{nakada96,kohmoto07}; these are not protected by
any topological symmetry since pristine graphene is gapless in the
bulk. On the other hand, a zigzag edge lying between graphene with
SOC and vacuum is known to have edge states which lie in the bulk
gap~\cite{kane05}; these states are topologically protected since
graphene with SOC of the Kane-Mele type has a non-zero Chern number
for each component of the spin. (A detailed discussion of edge
states in graphene can be found in Ref.~\onlinecite{lado15}).

A system consisting of pristine graphene separated from graphene
with SOC by a zigzag edge is gapless on one side. Hence if there are
states on the edge, they are not expected to be topologically
protected. However we will see below that for a given momentum along
the edge, these states lie in the gap of the bulk states, which means they
have the same momentum in both pristine graphene and graphene with SOC.
Hence these edge states cannot mix with the bulk states under
perturbations which conserve the momentum.

\begin{figure}[htb] \ig[width=2.6in]{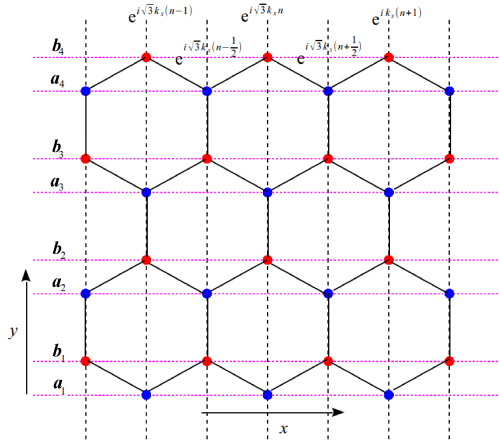}
\caption[]{Picture of the lattice used to calculate the
spectrum shown in Figs.~\ref{fig04} (a-d). The plane wave factors depend on
the momentum $k_x$. The label $n$ increases horizontally in the positive $x$
direction while the subscripts of $a_m$ and $b_m$ increase vertically in the
positive $y$ direction.} \label{fig03} \end{figure}

We are interested in studying states which are localized along an infinitely
long zigzag edge which runs along the $x$ direction. The momentum $k_x$
along the edge is therefore a good quantum number. (We have absorbed the
lattice spacing $d$ in the definition of $k_x$; hence $k_x$ is dimensionless).
We will denote the wave functions as $A_{m,n}$ and $B_{m,n}$, where the
coordinate $m$ increases vertically in the $y$ direction and the coordinate
$n$ increases horizontally in the $x$ direction. We assume that the wave
function is a plane wave in the $x$ direction so that $A_{m,n} = a_m
e^{i\sqrt{3}k_x n}$ or $a_m e^{i\sqrt{3}k_x (n+1/2)}$ depending on whether
$m$ is odd or even, and $B_{m,n} = b_m e^{i\sqrt{3}k_x n}$ or $b_m
e^{i\sqrt{3}k_x (n+1/2)}$ depending on whether $m$ is even or odd; this is
shown in Fig.~\ref{fig03}. We then obtain the equations (with $\ga = 1$)
\bea && - ~[2 \cos (\frac{\sqrt{3} k_x}{2}) b_m ~+~ b_{m-1}] \non \\
&& - 2 t_2 s^z ~[\sin (\sqrt{3} k_x) a_m ~-~ \sin (\frac{\sqrt{3} k_x}{2})
(a_{m-1} + a_{m+1})] \non \\
&& =~ E ~a_m, \non \\
&& - ~[2 \cos (\frac{\sqrt{3} k_x}{2}) a_m ~+~ a_{m+1}] \non \\
&& + 2 t_2 s^z ~[\sin (\sqrt{3} k_x) b_m ~-~ \sin (\frac{\sqrt{3} k_x}{2})
(b_{m-1} + b_{m+1})] \non \\
&& =~ E ~b_m, \label{eom} \eea
where we have taken into account the spin of the electron $s^z$.
Eqs.~\eqref{eom} imply that we effectively have a one-dimensional system
in which the site label $m$ goes from $-N_y/2$ to $N_y/2 -1$ for a finite
system with $2N_y$ sites (i.e., $N_y$ unit cells). Eqs.~\eqref{eom} will
give the energy $E$ as a function of the momentum $k_x$.

Eqs.~\eqref{eom} remain invariant under the following sets of transformations.

\noi (i) $k_x \to k_x + 2\pi/\sqrt{3}$, $a_m \to - (-1)^m a_m$, and
$b_m \to (-1)^m b_m$.

\noi (ii) $k_x \to - k_x$, and $s^z \to - s^z$.

\noi (iii) $k_x \to 2\pi/\sqrt{3} - k_x$, $E \to - E$, $a_m \to (-1)^m a_m$,
and $b_m \to (-1)^m b_m$.

\noi Using the above transformations and combinations of them we can
understand all the symmetries of the spectra shown in Figs.~\ref{fig04}
(a-d) below. [The transformations in (i) have a simple interpretation. The
solutions of Eqs.~\eqref{eom} must remain invariant if the momentum is
changed from $\vk$ to $\vk + {\vec G}_i$, where ${\vec G}_i$ is one of
the reciprocal lattice vectors given in Sec.~\ref{sec:intro}.
Since the $x$ component of both the ${\vec G}_i$ is equal to $2\pi/\sqrt{3}$,
we see that Eqs.~\eqref{eom} must remain the same under $k_x \to k_x +
2\pi/\sqrt{3}$].

We have numerically
solved Eqs.~\eqref{eom} for a system in which the upper half has SOC
while the lower half does not; more precisely, there is a SOC between sites
$a_l$ and $a_m$ (and between $b_l$ and $b_m$) only if $l, ~m$ are both larger
than zero and $|l-m| \le 1$.
The dividing line between pristine graphene and graphene with
SOC is therefore given by the zigzag edge consisting of the sites $A_{0,n}$
and $B_{0,n}$ where $n$ goes from $-\infty$ to $\infty$. The results are
presented in Figs.~\ref{fig04} (c-d) taking $t_2 = 0.1$. For comparison, we
show in Fig.~\ref{fig04} (a) the spectrum for pristine graphene; apart from
the gapless bulk states which are shaded blue (they almost form a continuum
since the bulk momentum $k_y$ takes a large number of almost continuous
values if $N_y$ is large), we see edge states which lie at exactly zero
energy between the values of $k_x = -4\pi/(3\sqrt{3})$ and $-2\pi/(3\sqrt{3})$
and between $k_x = 2\pi/(3\sqrt{3})$ and $4\pi/(3\sqrt{3})$.
Similarly, Fig.~\ref{fig04} (b) shows the spectrum for
graphene with SOC, taking $t_2 = 0.1$. The bulk spectrum is now gapped
at the Dirac points; the gap is given by $2 |\De_{so}| = 6 \sqrt{3} |t_2|$.
We notice four edge states which go between the lower and upper bands,
crossing zero energy at $k_x = \pi/\sqrt{3}$ (see Fig. 1 in
Ref.~\onlinecite{kane05}). If we now look at
the spectrum shown in Fig.~\ref{fig04} (c) for a system with a zigzag edge
lying between pristine graphene and graphene with SOC, we see that all the
states present in Figs.~\ref{fig04} (a-b) are also present here; in addition,
an extra set of edge states appear which lie very close to $E=\pm 1$ and
$k_x = \pm \pi/\sqrt{3}$. These are shown more clearly in Fig.~\ref{fig04} (d)
which is a zoomed in view of the region around $k_x = \pi/\sqrt{3}$.

\begin{widetext} \begin{center} \begin{figure}[h!]
\subfigure[]{\ig[width=2.6in]{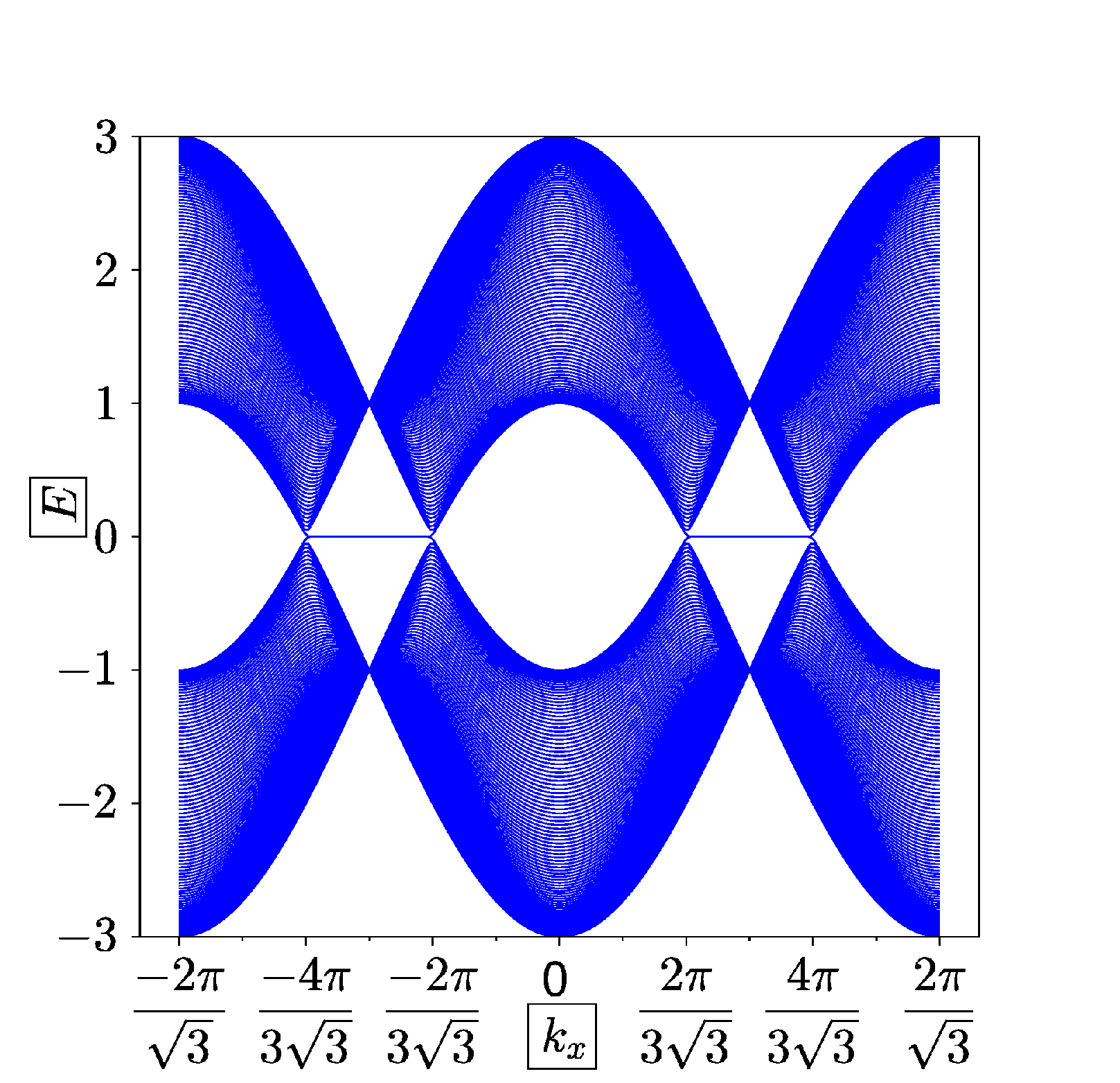}}
\subfigure[]{\ig[width=2.6in]{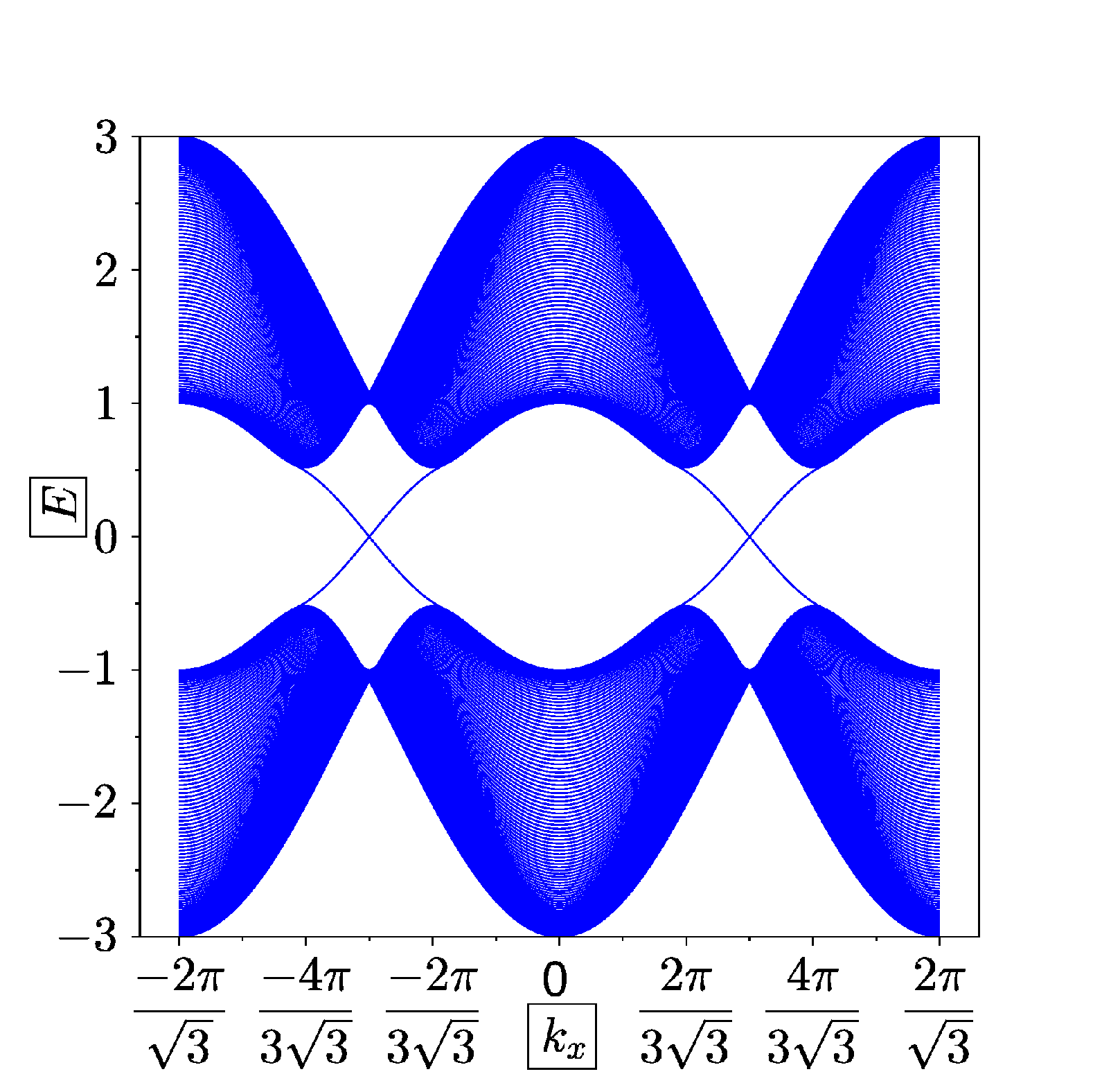}} \\
\subfigure[]{\ig[width=2.6in]{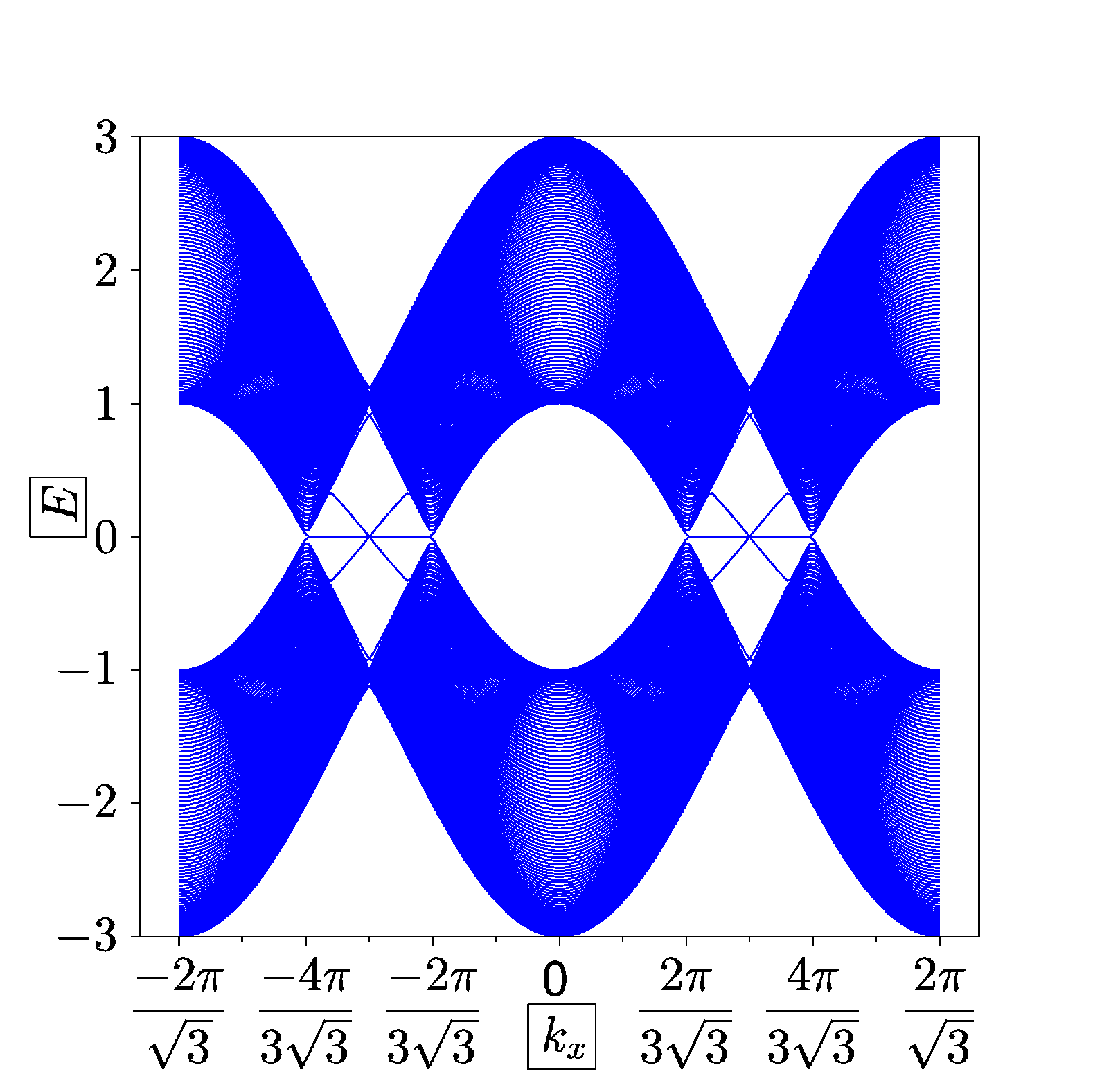}}
\subfigure[]{\ig[width=2.6in]{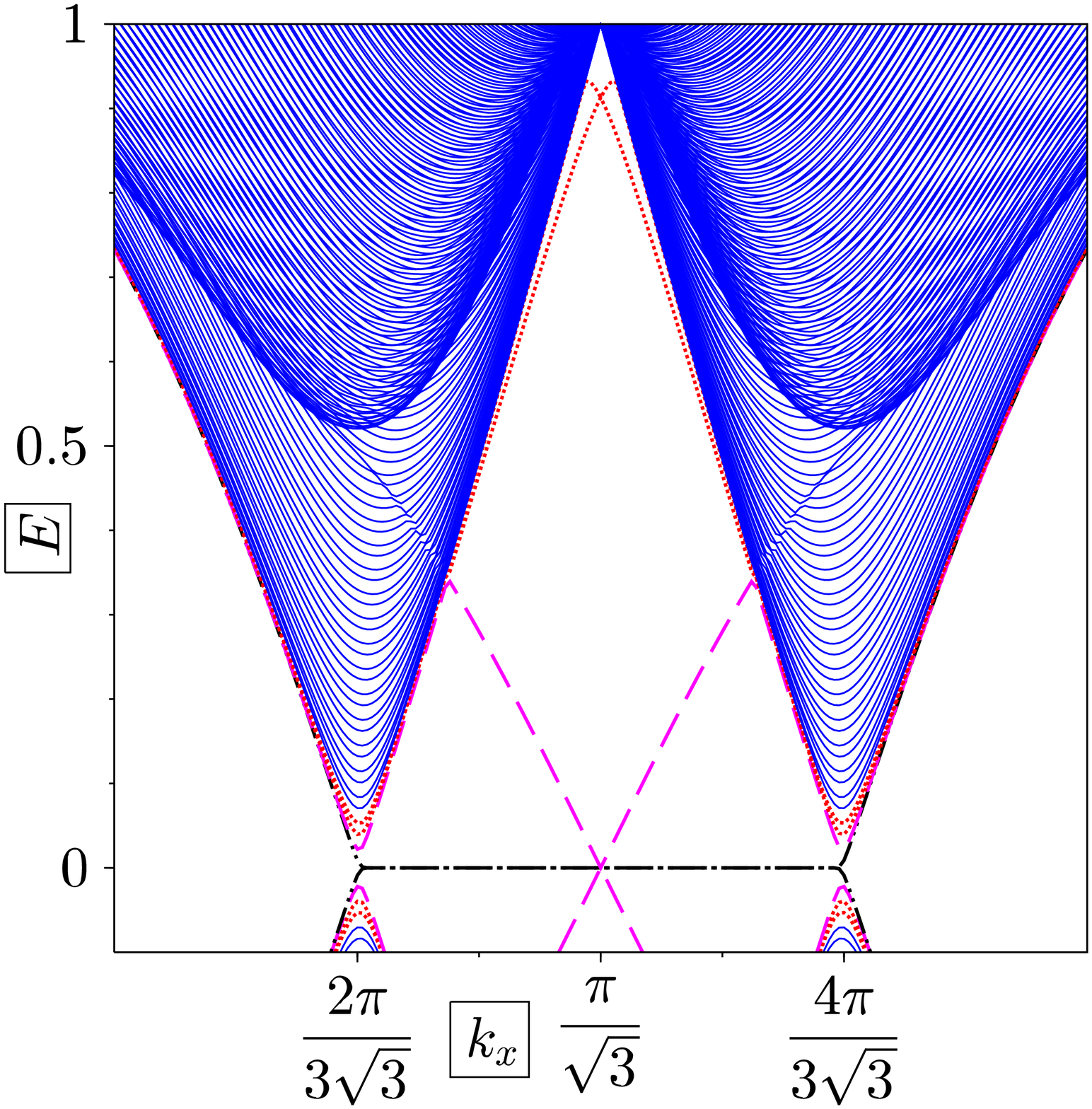}}
\caption[]{Energy-momentum dispersion for both spins ($s^z =
\pm 1$) for three systems:
(a) pristine graphene with a zigzag edge, (b) graphene with a SOC of strength
$t_2 = 0.1$ and a zigzag edge, and (c) a system with pristine graphene and
graphene with SOC ($t_2 = 0.1$) which are separated by a zigzag boundary. In
all the figures, $k_x$ denotes the momentum in the direction along the edge,
and the shaded regions denote the bulk states which form a continuum due to
the momentum $k_y$ in the direction transverse to the edge. ($E$ and $k_x$ 
are shown in units of $\ga$ and $1/d$ respectively). Three types of
edge states are visible: (i) edge states of pristine graphene in Figs. (a)
and (c) (these have $E=0$ for a range of $k_x$), (ii) edge states of graphene
with SOC in Figs. (b) and (c) (near $E=0$ these have $E$ varying linearly with
$|k_x \pm \pi/\sqrt{3}|$), and (iii) boundary states between pristine graphene
and graphene with SOC in Fig. (c) (these lie only slightly inside the gap).
Fig. (d) is a zoomed in view of the region around $k_x = \pi/\sqrt{3}$ and
positive energy which shows more clearly all the edge states; the black
dashed-dotted line shows the edge states of pristine graphene, the magenta
dashed lines show the edge states of graphene with SOC, and the red dotted
lines near the top show the boundary states between pristine graphene and
graphene with SOC.} \label{fig04} \end{figure} \end{center} \end{widetext}

\begin{figure}[htb] \ig[width=3.4in]{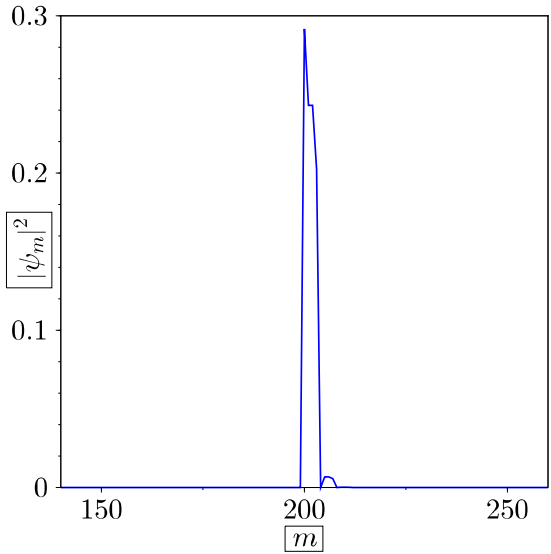}
\caption[]{$|\psi_m|^2$ (denoting $|a_m|^2$ and $|b_m|^2$
alternately) vs the coordinate $m$ of the edge state with momentum
$k_x = \pi/\sqrt{3}$ (in units of $1/d$) lying at the junction between 
pristine graphene and graphene with a SOC of strength $t_2 = 0.1$ (in units of
$\ga$).} \label{fig41} \end{figure}

We have analytically studied these edge states lying between pristine
graphene and graphene with SOC. We find that they take a particularly simple
form if $k_x = \pm \pi/\sqrt{3}$; in this case, Eqs.~\eqref{eom} reduce to
\bea \pm 2 t_2 s^z (a_{m+1}+a_{m-1}) -b_{m-1} &=& E ~a_m, \non \\
\pm 2 t_2 s^z (b_{m+1}+b_{m-1}) -a_{m+1}&=& E ~b_m, \label{eomsim} \eea
These equations admit a solution
\bea E_0 &=& \pm 1/\sqrt{1+2|t_2|}, \non \\
a_m, ~b_m &\sim& \left( \frac{|2t_2|}{1 + 2 |t_2|} \right)^{[m/2]}
~~{\rm for}~~ m \ge 0, \label{enexp} \eea where $[m/2]$ denotes the
largest integer less than or equal to $m/2$, and $a_m = b_m = 0$ for
$m < 0$. (In the limit $t_2 \to 0$, the wave function remains
non-zero only on the four sites $a_0$, $b_0$, $a_1$, and $b_1$).
Thus the energy lies within the bulk gap on the side of graphene
with SOC and the wave function decays exponentially on the side of
graphene with SOC and is exactly zero on the pristine graphene side;
this is shown in Fig.~\ref{fig41} for $t_2 = 0.1$. For small $t_2$,
Eq.~\eqref{enexp} shows that the wave function decays as
$|2t_2|^{[m/2]}$ at a site which is $m$ unit cells away from the
junction inside the region of graphene with SOC; this implies that
the decay length is proportional to $-3d/[\ln (2|t_2|/ \ga)]$,
where we have restored all the dimensionful parameters. Thus the
decay length goes to zero as $t_2 \to 0$. Note that this behavior is
in complete contrast with those of conventional edge states where
the decay length diverges as the bulk energy gap vanishes. This
indicates that while conventional edge states delocalize and merge
with the continuum bulk states in the limit of vanishing gap, edge
states at the boundary of graphene with SOC and pristine graphene
become completely localized at the zigzag edge separating the two
regimes. Away from the special values of $k_x = \pm \pi/\sqrt{3}$,
it is difficult to obtain analytical solutions for the edge states.
However, we find numerically that for small $t_2$, the edge states
exist only in a small range of values of $k_x$ close to $\pm
\pi/\sqrt{3}$. The decay length of these states grows as $k_x$
approaches the ends of its allowed range beyond which the edge
states merge with the continuum of bulk states.

It may seem surprising that such localized edge
states exists even if $t_2 = 0$ when the system has pristine
graphene everywhere. This behavior becomes obvious from
Eq.~\eqref{eomsim} which admits solutions with $E=\pm 1$ and $a_m = \pm
b_{m-1}$ for any value of $m$ in this limit. Further, the presence
of a state along a zigzag edge for any value of $m$ suggests an
unusually large number of states, increasing linearly with $N_y$, at
$E = \pm 1$; this is consistent with the Van Hove singularity in the
density of states of pristine graphene at those two energies~\cite{neto09}.

Finally, we have numerically studied the fate of the edge states for an
armchair edge. We have found that states do not appear at an armchair 
edge lying between pristine graphene and graphene with SOC.

We note that while graphene with SOC of the Kane-Mele type is a 
topological system with a bulk gap, pristine graphene is gapless in the bulk 
and is not a topological system. Due to the gapless nature of pristine 
graphene, an edge shared between it and graphene with SOC is different from an
edge between topologically trivial and non-trivial insulators. In particular,
the former need not host localized edge states since any low-energy states
may be delocalized since bulk pristine graphene is gapless. Thus
one does not need to have gapless localized states on such a boundary;
however, our work shows that a zigzag edge between graphene with SOC and
pristine graphene has such states while an armchair edge does not. We
note that these edge states do not have topological protection in contrast
to their counterparts at edges separating topologically trivial 
and non-trivial insulators.

It is interesting to compare our results on edge states 
between graphene with SOC and pristine graphene to the 
edge states which appear between (i) graphene with SOC and vacuum, and 
(ii) between pristine graphene and vacuum. These two cases have been 
studied extensively in the literature. It is known that edge states appear 
at both zigzag and armchair edges between graphene with SOC and 
vacuum~\cite{kane05,sengupta06,jiang11}. This is because graphene with SOC 
is a topological system; hence states appear on any edge (zigzag or armchair) 
between this system and the vacuum, and all these edge states are 
topologically protected. On the other hand, pristine graphene is a 
gapless and non-topological system; hence its edges with any other system may 
or may not host any states. It turns out that a zigzag edge between pristine 
graphene and vacuum has edge states, but an armchair edge between pristine 
graphene and vacuum does not host any states~\cite{nakada96,kohmoto07,lado15}.

We observe that Fig.~\ref{fig04} shows states at a zigzag edge for all the
three cases discussed above. Figure \ref{fig04} (a) shows edge states between
pristine graphene and vacuum; these are dispersionless and lie exactly at zero 
energy. Figure \ref{fig04} (b) shows edge states between graphene with SOC 
and pristine graphene; these have a dispersion and go through zero energy at 
two particular momenta. Finally, the top part of Fig.~\ref{fig04} (d) shows 
edge states between graphene with SOC and pristine graphene; these appear 
only in a small range of momentum and have a dispersion which lies close 
to that of the bulk states.

\section{Effect of localized impurity}
\label{sec:imp}

In this section, we will study the effect that an impurity placed
somewhere in graphene has on the LDOS $\rho ({\vec r},E)$ as a
function of the position $\vec r$ and energy $E$. By the LDOS we
will mean the sum of the densities on the $a$ and $b$ sites at the
unit cell labeled $\vec r$; we will also sum over the electron spin.
For pristine graphene this problem was studied in
Refs.~\onlinecite{cheianov06,mariani07,bena08,bena09}; our aim is to
go beyond those papers by studying additional characteristics in
LDOS due to the presence of the SO term and/or Zeeman field. In what
follows, we shall carry out an analysis of the LDOS in the weak
impurity potential regime where perturbation theory holds.

To compute the LDOS in this regime, we can use the standard $T$-matrix
formalism developed for pristine graphene in Refs.~\onlinecite{bena08} and
\onlinecite{bena09}. In the absence of any impurities, the density of states
is given by
\bea \rho_0 ({\vec r}, E) &=& -~\frac{1}{\pi} ~Im ~\{ tr ~[G_0 ({\vec r},
E)]\}, \non \\
G_0 ({\vec r}, E) &=& \la {\vec r} | ~\frac{1}{E ~-~ H_0 ~+~ i \ep}
~| {\vec r} \ra, \label{ldos1} \eea
where the Green's function $G_0 ({\vec r}, E)$ is a $4 \times 4$ matrix in
sublattice and spin space, and $\ep$ is an infinitesimal positive number.
Note that the LDOS in Eq.~\eqref{ldos1} is independent of $\vec r$ as
a consequence of the translation symmetry of the system in the
absence of impurities. In the presence of impurities, the total
Hamiltonian is given by $H_0 + V_{\rm imp}$ and the LDOS is given by
\bea \rho ({\vec r}, E) &=& -~\frac{1}{\pi} ~Im ~\{ tr ~[G ({\vec
r}, E)] \}, \non \\
G ({\vec r}, E) &=& \la {\vec r} | ~\frac{1}{E ~-~ H_0 ~-~ V_{imp}
~+~ i \ep} ~| {\vec r} \ra. \label{ldos3} \eea
We now consider an impurity of strength $u$ which is placed at the $a$ site
of a unit cell located at ${\vec r}_0$ with a potential given by
\beq V_{imp} ~=~ \sum_{\al = \ua,\da}~ u ~a^\dg_{{\vec r}_0,\al} ~
a_{{\vec r}_0,\al}. \label{vimp} \eeq
Within the $T$-matrix formalism and to first order in perturbation theory,
the change in the LDOS due to the impurity is given by \bea \de \rho
({\vec r}, E) &=& -~\frac{1}{\pi} ~Im ~tr ~[ \la {\vec r} |
\frac{1}{E ~-~ H_0 ~+~ i \ep} ~v_{imp} ({\vec r}_0) \non \\
&& ~~~~~~~~~\times ~\frac{1}{E ~-~ H_0 ~+~ i \ep} |{\vec r} \ra ],
\label{ldos4} \eea
We define the two-point real space Green's function for graphene without
impurities as
\bea {\cal G}_0 ({\vec r}_1, {\vec r}_2, E) &=& \la {\vec r}_1 | ~
\frac{1}{E ~-~ H_0 ~+~ i \ep} ~| {\vec r}_2 \ra \non \\
&=& \int \frac{d^2 k}{(2\pi)^2} \sum_\al \frac{u_{\vk,\al}
u_{\vk,\al}^\dg ~e^{i \vk \cdot ({\vec r}_1 - {\vec r}_2)}}{E ~-~
E_{\vk,\al} + i \ep}. \non \\
&& \label{twopointg} \eea
One can then write the change in the LDOS as
\beq \de \rho ({\vec r}, E) = -\frac{1}{\pi} ~Im ~\{ tr ~[ {\cal G}_0
({\vec r}, {\vec r}_0, E) ~v_{imp} ({\vec r}_0) ~{\cal G}_0 ({\vec r}_0,
{\vec r}, E)] \}. \label{delrho} \eeq
While one can numerically compute $\de \rho ({\vec r}, E)$ using
Eqs.~\eqref{twopointg} and \eqref{delrho}, the results seem to depend
sensitively on the values of the momentum spacing $\De k_x, ~\De k_y$
and $\ep$ that one chooses.

In order to avoid such cutoff dependences, we have directly
computed $\de \rho ({\vec r}, E)$ by numerically calculating $\rho
({\vec r}, E)$ with and without the impurity and then taking the
difference. The calculations are carried out as follows.
We consider a lattice in which the integers $n_1, ~n_2$ go from $1$
to some integer $N$; hence the lattice has $N^2$ unit cells and
$2N^2$ sites. We impose periodic boundary conditions. (There are two 
reasons for choosing such a boundary condition. First, it ensures
that momentum is a good quantum number in the absence of an impurity.
Second, we will study below the Fourier transform of the change in 
the local density of states produced by the impurity; this requires
periodic boundary conditions in order to define a momentum). 
Corresponding to $n_1$ and $n_2$,
we define two momenta $l_1$ and $l_2$ each of which goes from $-\pi$
to $\pi - 2\pi/N$ in steps of $2\pi/N$ just as we expect for
orthogonal Cartesian coordinates. In terms of the quantities $n_i$
and $l_i$, the phase of plane waves is given by $l_1 n_1 + l_2 n_2$.
We now go to the non-orthogonal position and momentum vectors of the
hexagonal lattice by observing that the real space position on the
lattice and the momentum $\vk = (k_x,k_y)$ will satisfy $l_1 n_1 +
l_2 n_2 = k_x n_x + k_y n_y$ provided that $k_x = l_1 /(\sqrt{3} d)$ and
$k_y = (2l_2 - l_1)/(3d)$.
Given the ranges of of $l_1, ~l_2$ stated
above, we see that in the limit $N \to \infty$, the Brillouin zone
will be a rhombus with corners at $(\pi /d) (-1/\sqrt{3}, -1/3)$,
$(\pi /d) (1/\sqrt{3},-1)$, $(\pi /d) (1/\sqrt{3}, 1/3)$, and $(\pi
/d) (-1/\sqrt{3}, 1)$; the area of the rhombus is $8 \pi^2/(3 \sqrt{3} d^2)$.

\begin{widetext} \begin{center} \begin{figure}[htb]
\subfigure[]{\ig[width=3.0in]{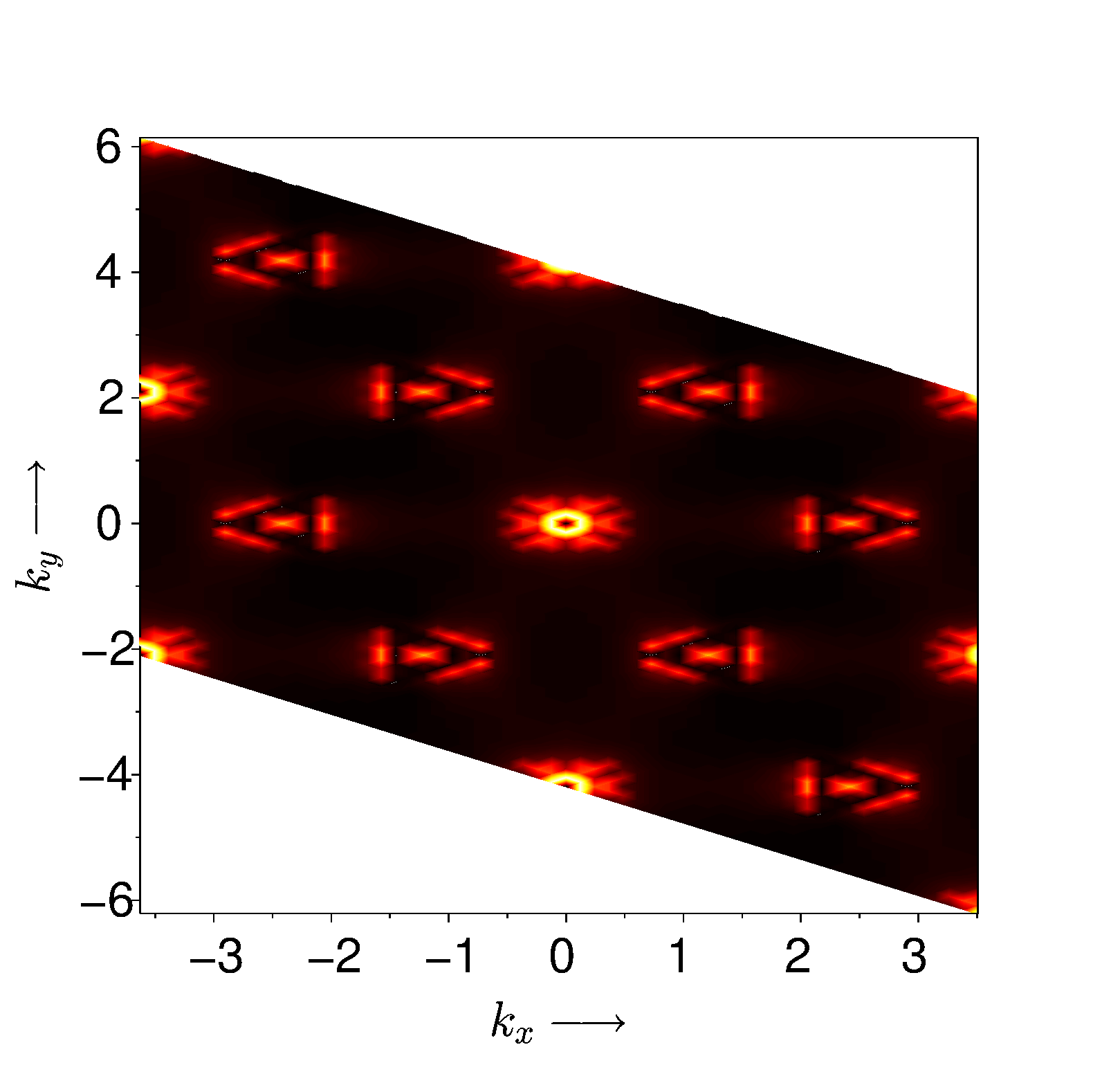}}
\subfigure[]{\ig[width=3.0in]{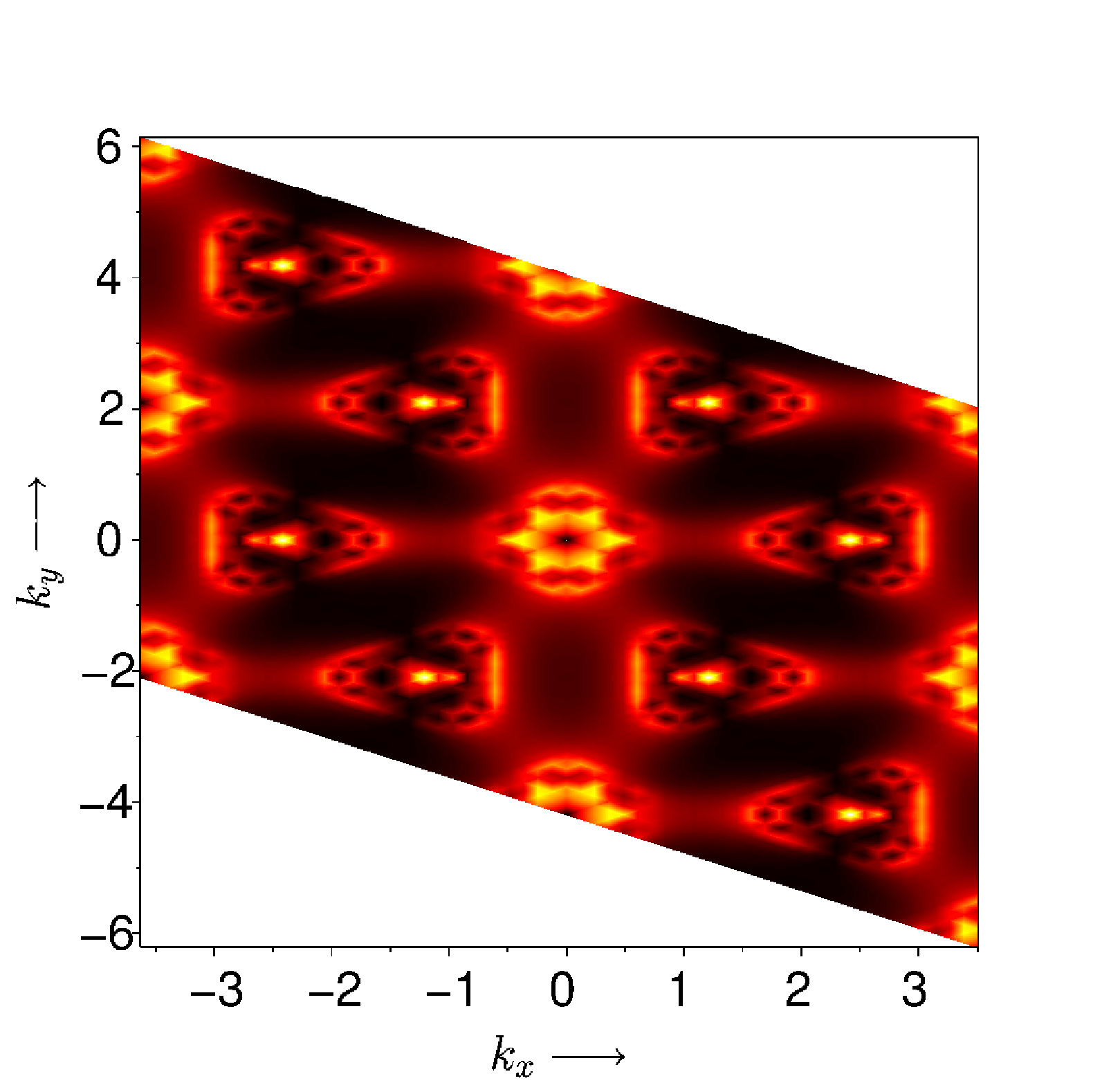}}
\subfigure[]{\ig[width=0.5in]{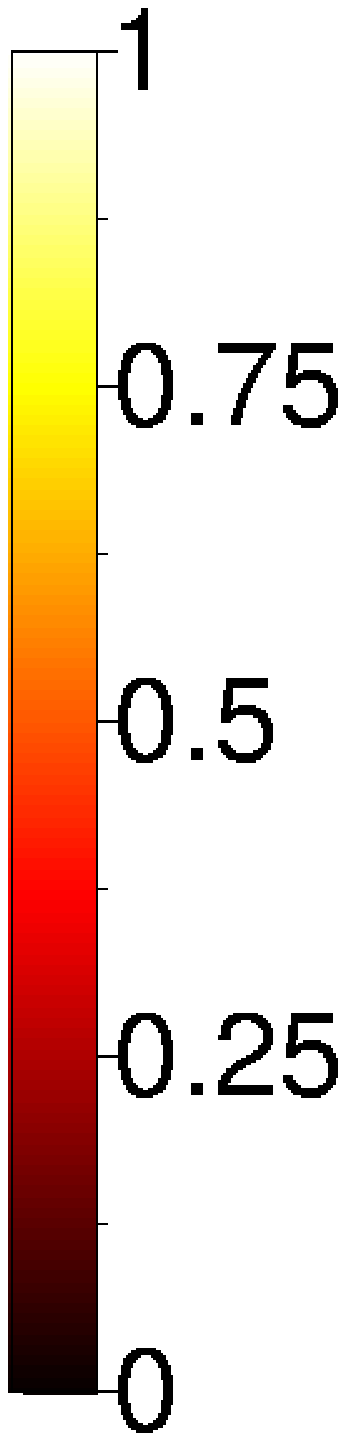}}
\caption{Fourier transform of change in LDOS at $E ~=~ 0.382$
and $0.502$ when an impurity is placed on the $a$ site in one particular unit
cell in pristine graphene (no SOC and no Zeeman field). The
calculation has been done on a $30 \times 30$ lattice. (The impurity strength
$u = 0.1$ has been divided out). The actual minimum and maximum values of the
LDOS are $(0,1.02)$ and $(0,0.58)$ respectively. The area of each picture is
four times the Brillouin zone area. ($E$ and $u$ are in units of $\ga$,
while $k_x$ and $k_y$ are in units of $1/d$).} \label{fig05} \end{figure} 
\end{center} \end{widetext}

\begin{widetext} \begin{center} \begin{figure}[htb]
\subfigure[]{\ig[width=3.0in]{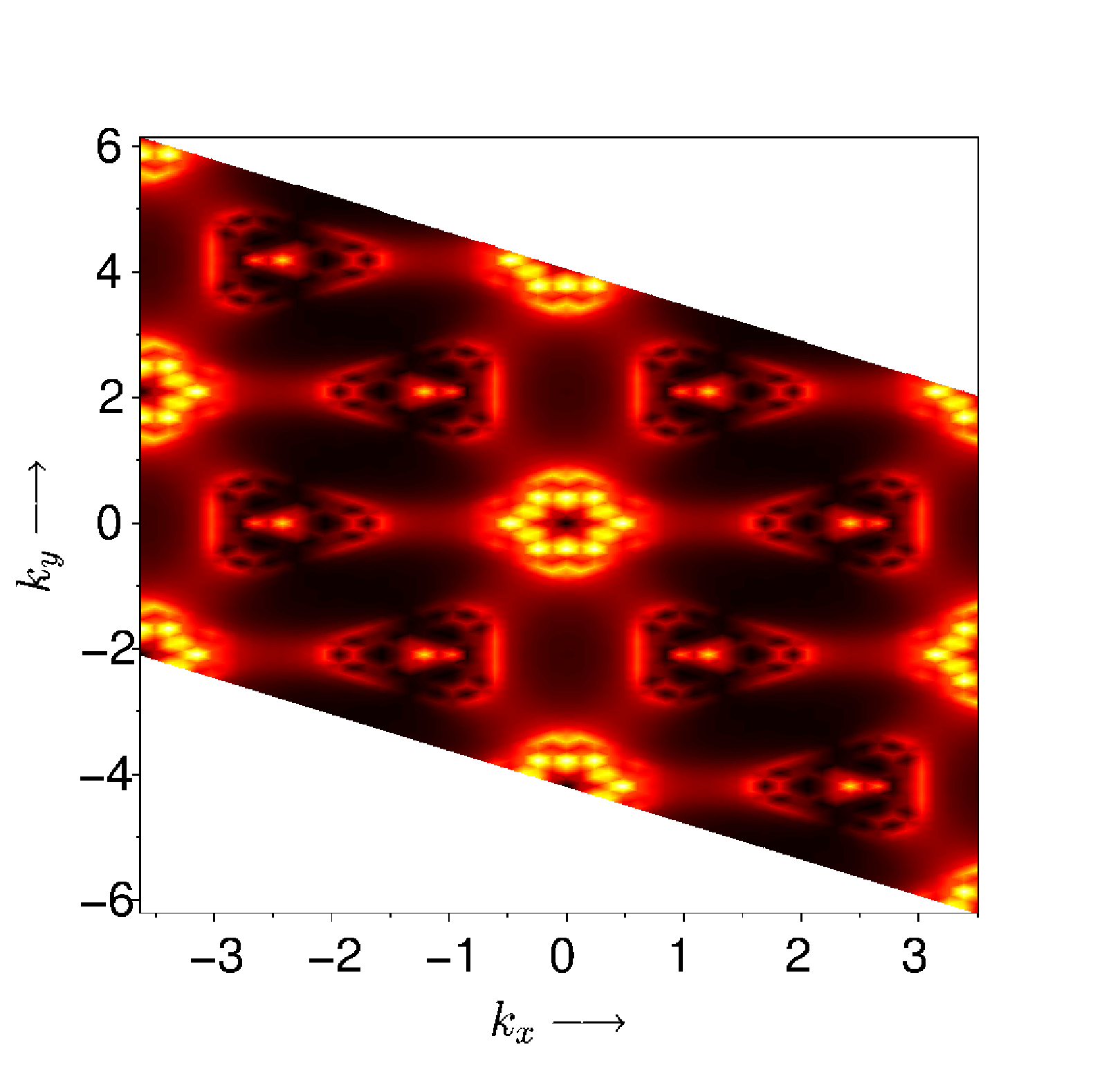}}
\subfigure[]{\ig[width=3.0in]{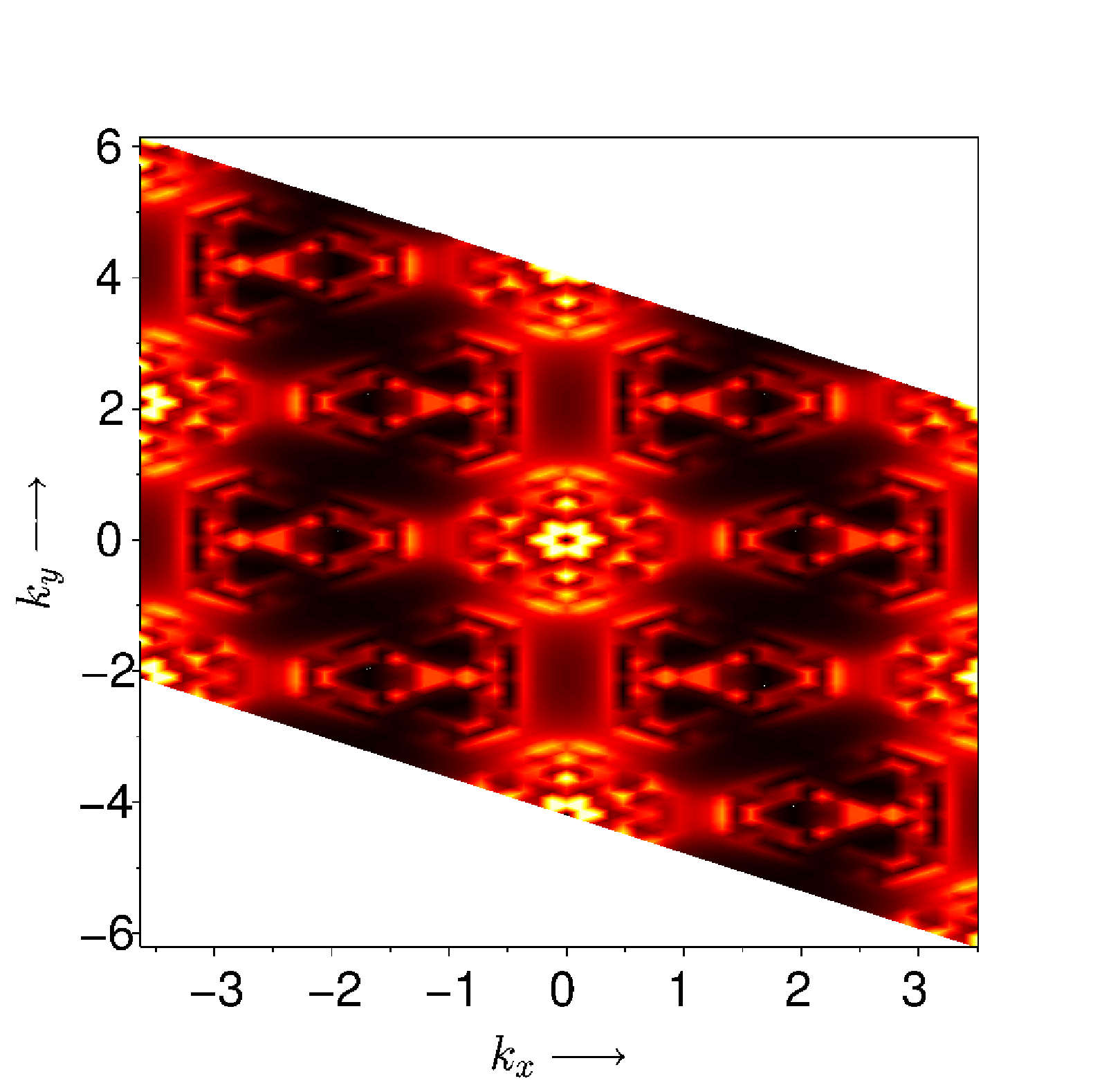}}
\subfigure[]{\ig[width=0.5in]{grafig05c.ps}}
\caption{Fourier transform of change in LDOS at $E=0.382$ and
$0.517$ when an impurity is placed on the $a$ site in one particular unit 
cell, with a SOC of strength $t_2 = 0.05$ and a Zeeman field $b_x = 0.2$. 
The calculation has been done on a $30 \times 30$ lattice. (The impurity 
strength $u = 0.1$ has been divided out). The actual minimum and maximum 
values of the LDOS are $(0,1.20)$ and $(0,1.58)$ respectively. The area of 
each picture is four times the Brillouin zone area. ($E, ~t_2, ~b_x$ and $u$
are in units of $\ga$, while $k_x$ and $k_y$ are 
in units of $1/d$).} \label{fig06} \end{figure} \end{center} \end{widetext}

\begin{widetext} \begin{center} \begin{figure}[htb]
\subfigure[]{\ig[width=3.0in]{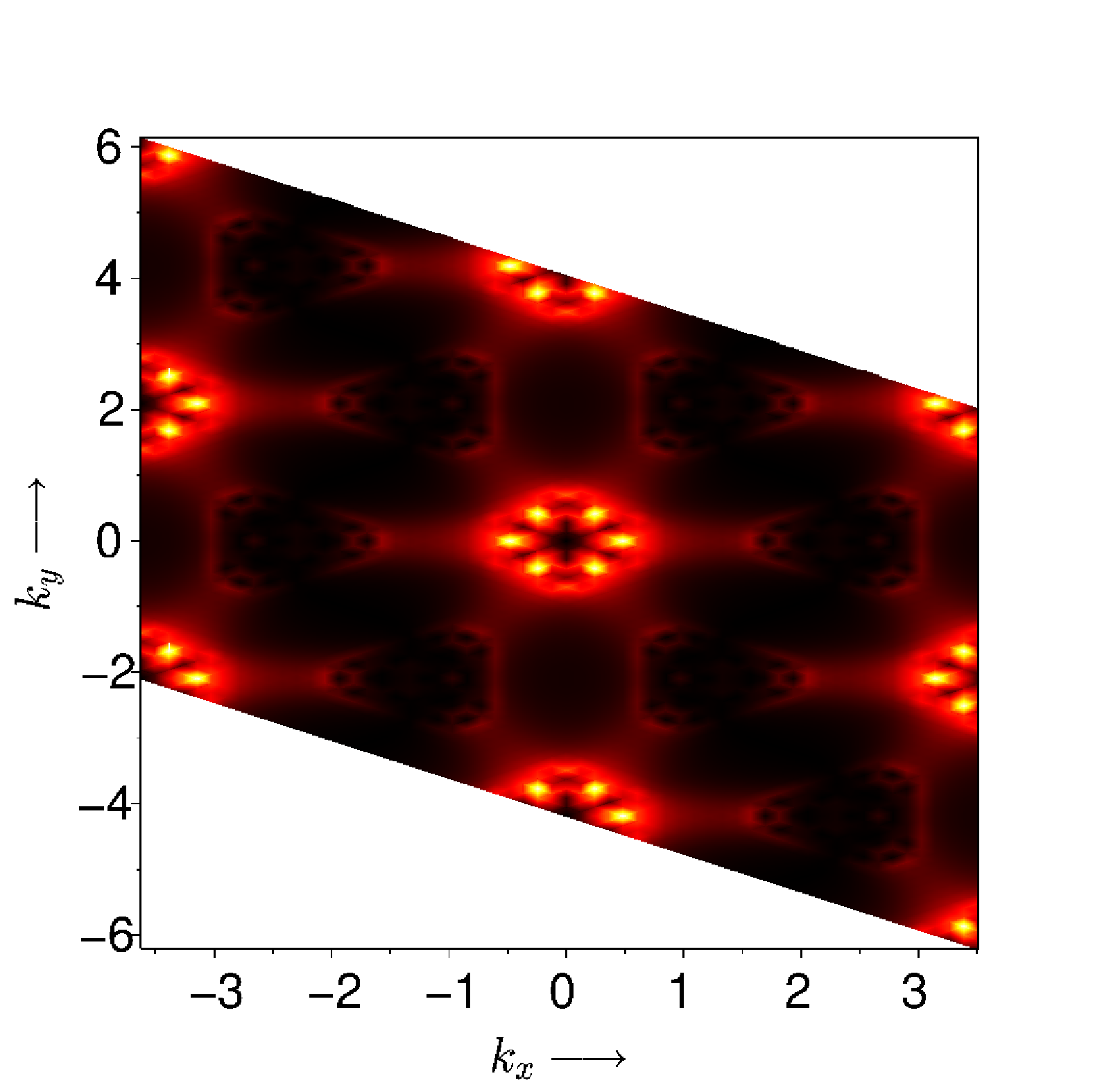}}
\subfigure[]{\ig[width=3.0in]{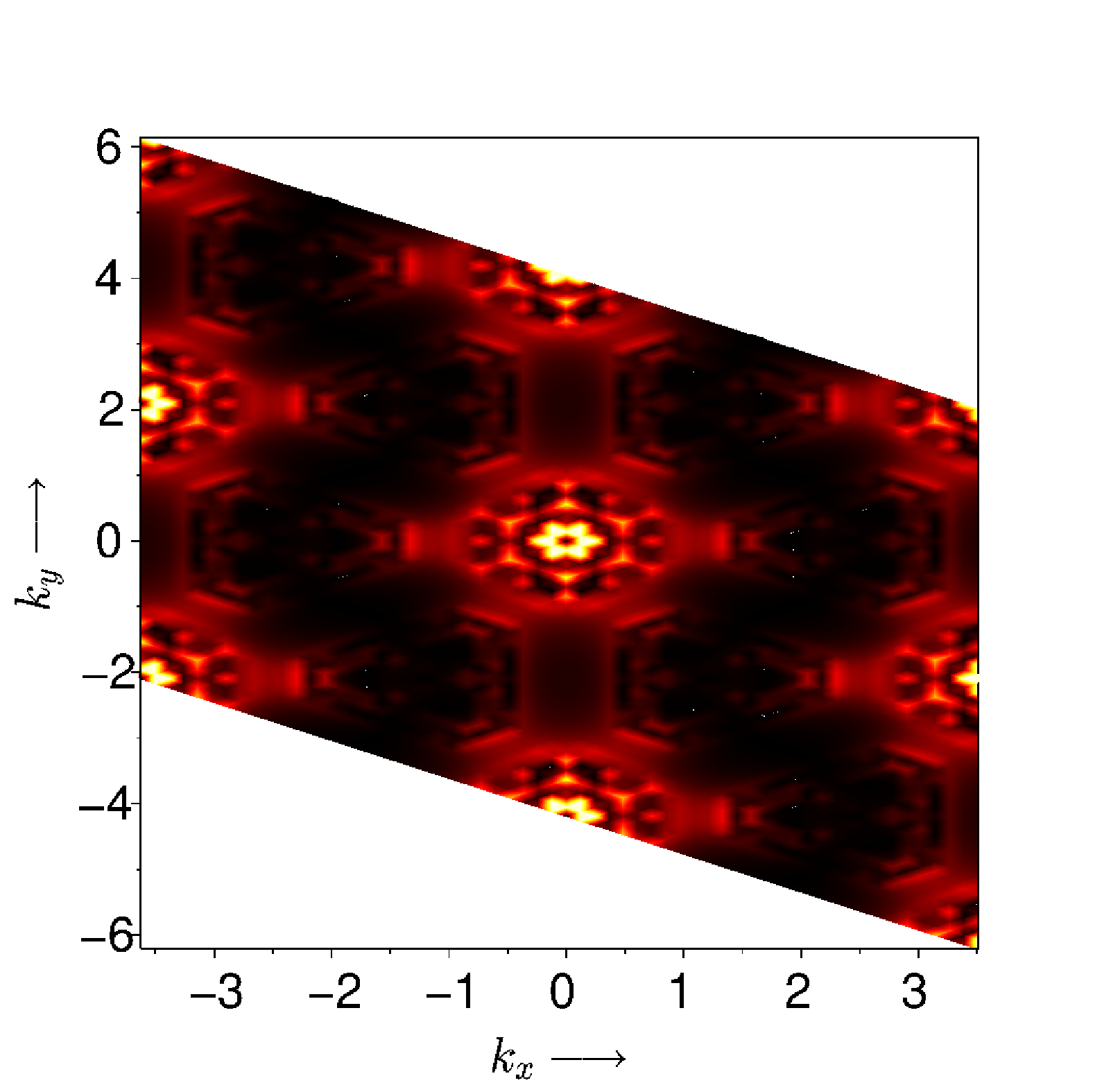}}
\subfigure[]{\ig[width=0.5in]{grafig05c.ps}}
\caption{Fourier transform of change in LDOS at $E=0.382$ and
$0.517$ when an impurity of strength $0.1/6$ is placed at each of the six
sites around one particular hexagon (hence the integrated strength is equal
to $0.1$). There is a SOC of strength $t_2 = 0.05$ and a Zeeman field $b_x = 
0.2$. The calculation has been done on a $30 \times 30$ lattice. (The impurity 
strength $u = 0.1$ has been divided out). The actual minimum and maximum 
values of the LDOS are $(0,0.03)$ and $(0,0.05)$ respectively. The area of 
each picture is four times the Brillouin zone area. ($E, ~t_2, ~b_x$ and $u$
are in units of $\ga$, while $k_x$ and $k_y$ are in units of $1/d$).} 
\label{fig66} \end{figure} \end{center} \end{widetext}

\begin{widetext} \begin{center} \begin{figure}[htb]
\subfigure[]{\ig[width=3.0in]{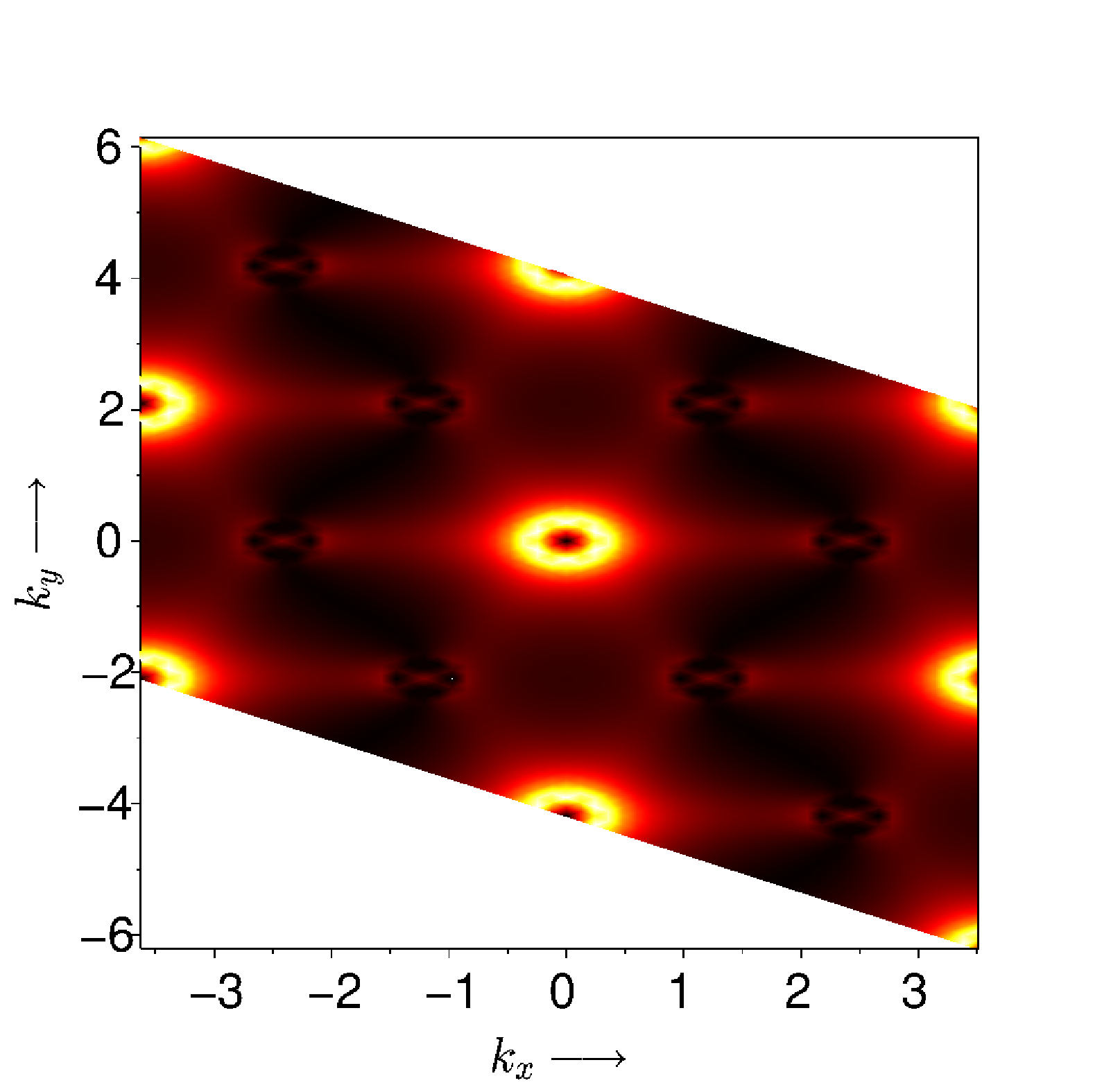}}
\subfigure[]{\ig[width=3.0in]{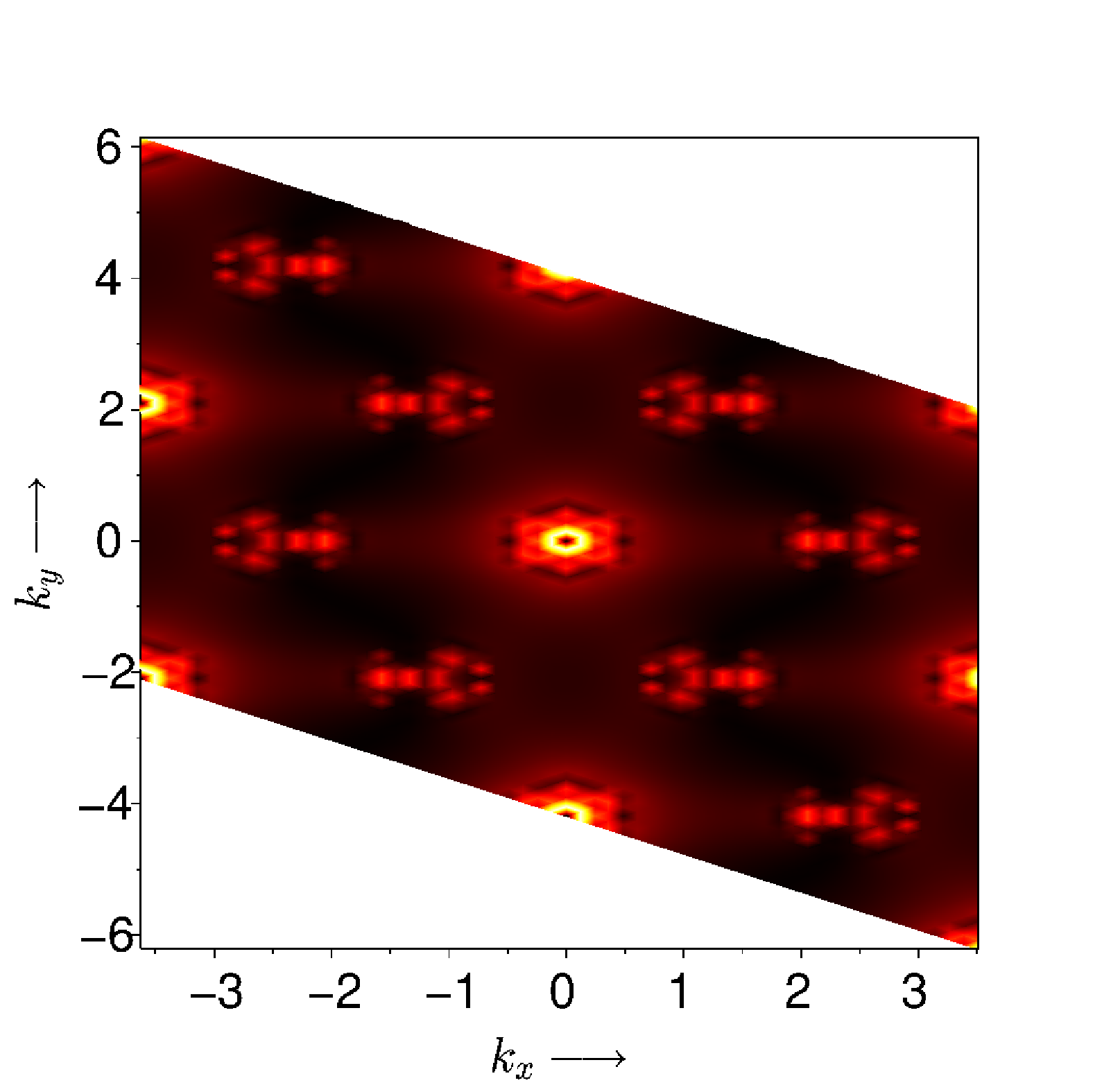}}
\subfigure[]{\ig[width=0.5in]{grafig05c.ps}}
\caption{Fourier transform of change in LDOS at $E=0.331$ and
$0.483$ when a magnetic impurity (coupling to $s^x = \pm 1$ with strengths
$\pm 0.1$ respectively) is placed on the $a$ site in one particular unit cell,
with a SOC of strength $t_2 = 0.05$ and no Zeeman field $b_x$. The calculation 
has been done on a $30 \times 30$ lattice. (The impurity strength $u = 0.1$ 
has been divided out). The actual minimum and maximum values of the LDOS are 
$(0,0.004)$ and $(0,0.007)$. The area of each picture is four times the 
Brillouin zone area. ($E, ~t_2, ~b_x$ and $u$ are in units of $\ga$, while
$k_x$ and $k_y$ are in units of $1/d$).} \label{fig67} \end{figure} 
\end{center} \end{widetext}

Since we are only interested in the change in the LDOS to first order in the
impurity strength $u$, we will take $u$ to be a small number and calculate
$\de \rho ({\vec r}, E) /u$. We first consider pristine graphene (i.e., in
the absence of SO coupling and Zeeman field) when an impurity of strength
$u=0.1$ is placed at the $a$ site of the unit cell centered at
${\vec r}_0 = (0,0)$.
The absolute value of the Fourier transform of $\de \rho ({\vec r}, E)/u$ is
shown in Fig.~\ref{fig05} for different values of $E$, with a $30 \times 30$
lattice (i.e., $N=30$). Since the energy spectrum is found to have an exact
or nearly exact six-fold degeneracy at most energies, we will calculate the
LDOS by summing over the contributions from the six states with energy closest
to the desired value of $E$. Figure \ref{fig05} shows that we get large and
sharp peaks at the Dirac points, namely, the six points forming a hexagon
around the center. [The six points are given by $(4\pi/3\sqrt{3}d)$ times
$(1,0)$, $(-1/2,\sqrt{3}/2)$ and $(-1/2,-\sqrt{3}/2)$, and $(4\pi/3\sqrt{3}d)$
times $(-1,0)$, $(1/2,\sqrt{3}/2)$ and $(1/2,-\sqrt{3}/2)$ which are
respectively equal to $\vec K$ and $\vec K'$ up to reciprocal lattice
vectors]. The peaks broaden as we move away from zero energy.
(We note that the normalization of the LDOS calculated in this way is
arbitrary to the extent that we have done the calculations for a particular
system size and have not normalized the results to take that into account).
We also note that the Fourier transform of $\de \rho ({\vec r}, E)$ is
always zero at $\vk = (0,0)$ since that is just the difference in the
number of states at that energy with and without the impurity, and we have
chosen $E$ in such a way that the impurity does not change that number.

In Fig.~\ref{fig06}, we show the absolute value of the Fourier
transform of $\de \rho ({\vec r}, E) /u$ when an impurity of
strength $u=0.1$ is placed at the $a$ site of the unit cell centered
at ${\vec r} = (0,0)$, when there is a SOC of strength $t_2 = 0.05$
and a Zeeman field $b_x = 0.2$. Comparing Figs.~\ref{fig05} and
\ref{fig06}, we find that the SOC and Zeeman field broaden the peaks
at the Dirac points. This is expected since the SOC and Zeeman field
open a gap and broaden the Dirac points in the dispersion shown in
Fig.~\ref{fig02}. Thus we find that the breadth of the LDOS peaks at
the Dirac point is a measure of the strength of the SOC and/or
Zeeman field in graphene.

Next, we study a distribution of non-interacting impurities instead
of a single fixed impurity studied earlier. More specifically, we
place impurities of strength $u_i=0.1/6$ on each of the six sites
around a graphene hexagon, so that the integrated impurity strength
is $u=\sum_{i=1,6} u_i= 0.1$ as before. We compute the absolute
value of the Fourier transform of $\de \rho ({\vec r}, E) /u$, as
shown in Fig.~\ref{fig66}, for $t_2 = 0.05$ and a Zeeman field $b_x
= 0.2$. Comparing Figs.~\ref{fig06} and \ref{fig66}, we see that
there no peaks at the Dirac points when the SOC is present. Further,
the maximum value of the Fourier transform is now much smaller than
in the case when the impurity is present only on a single site.
These observations can be understood as follows. The Fourier
transform of the two-point Green's function in Eq.~\eqref{twopointg}
is particularly large at the Dirac momenta ${\vec K}$ and ${\vec
K'}$. Equation \eqref{delrho} shows that $\de \rho ({\vec r}, E)$ is
composed of two such Green's functions. Hence the Fourier transform
of $\de \rho ({\vec r}, E)$ will be peaked at the difference of
${\vec K}$ and ${\vec K'}$, namely, at ${\vec K}$ and ${\vec K'}$
(since ${\vec K} - {\vec K'} = {\vec K'}$), provided that the
Fourier transform of $v_{imp}$ does not vanish at ${\vec K}$ and
${\vec K'}$. This is true if there is an impurity at a single site.
However, if there are impurities of equal strengths at the six sites
around a hexagon, the Fourier transform of this, given by
$\sum_{i=1}^6 e^{i \vk \cdot {\vec r}_i}$, vanishes at $\vk = {\vec
K}$ and ${\vec K'}$ due to destructive interference between
contribution from each point. Hence the Fourier transform of $\de
\rho ({\vec r}, E)$ is negligible at ${\vec K}$ and ${\vec
K'}$~\cite{bena08,bena09}. Such a cancellation is unique for Dirac
electrons in graphene and has been pointed out in the context
of LDOS~\cite{bena08} and STM spectra~\cite{saha10} of a single impurity
placed at the hexagon center in graphene; our work here points out
that such a cancellation is qualitatively important for understanding
the structure of LDOS for distributed impurities in graphene.

Finally, we study the LDOS in the presence of a magnetic impurity at a
single site of graphene with SOC. Such an impurity provides a direct
coupling to electron spin at that site. More specifically, we assume
that the impurity is on the $a$ site of a unit cell located at
${\vec r}_0$ and couples with strength $u$ to the $x$-component of
the spin: \beq V_{imp} ~=~ u ~(a^\dg_{{\vec r}_0,\ua} a_{{\vec
r}_0,\da} + a^\dg_{{\vec r}_0,\da} a_{{\vec r}_0,\ua}).
\label{magimp} \eeq For $u=0.1$ and a SOC of strength $t_2 = 0.05$,
the Fourier transform of the change in the LDOS is shown in
Fig.~\ref{fig67} for two values of the energy $E$. Comparing
Figs.~\ref{fig06} and \ref{fig67}, we find that the scale of the
change in the LDOS is much smaller for a magnetic impurity compared
to a non-magnetic impurity of the same strength, namely,
$0.004-0.007$ versus $1.20-1.58$. This can be partly understood as
follows. Since $s^z$ and $s^x$ anticommute, a unitary transformation
of the Hamiltonian by $s^z$ leaves the SOC parameter $t_2$ unchanged
but flips the impurity parameter $u \to - u$. Since the LDOS must be
invariant under this unitary transformation, it must be an even
function of $u$. To lowest order, therefore, the change in the LDOS
must be of order $u^2$ for a magnetic impurity, while it is of order
$u$ for a non-magnetic impurity. For $u=0.1$, we therefore expect
the change in the LDOS to be about 10 times smaller for a magnetic
impurity. Thus we find that a magnetic impurity will have a smaller
effect on LDOS compared to its non-magnetic counterpart.

\section{Spin active graphene junctions}
\label{sec:junc}

In this section we will study the differential conductance $G$
for either a junction of graphene with SOC and pristine graphene or
two regions of pristine graphene which are separated from each other
by finite width strips of various kinds, such as graphene with SOC
or in an external Zeeman field. In Sec.~\ref{sec:condan}, we carry
out an analytical calculation for the differential conductance from
a continuum theory. This will be followed by Sec.~\ref{sec:condnum},
where we will numerically calculate $G$ for finite-sized systems
using a lattice model. A comparison between the results obtained by
these two approaches is given in Sec.\ \ref{sec:spinactive}.

\subsection{Analytical calculation using continuum models}
\label{sec:condan}

\begin{figure}[htb] \ig[width=3.4in]{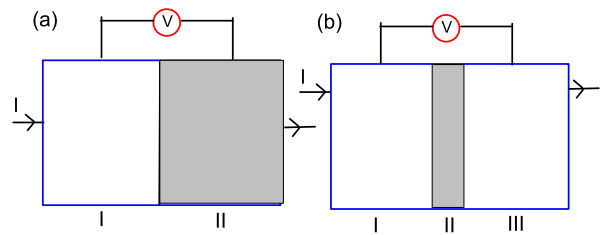}
\caption[]{Schematic representation of the junction geometry
studied in Sec.~\ref{sec:condan}. The shaded region represents graphene with
SOC while the white region denotes pristine graphene.} \label{fig:pic}
\end{figure}

In this section, we analyze transport in graphene junctions
with SOC. The geometry for such junctions which
will be studied in this section is shown in Fig.~\ref{fig:pic}. We
begin with an analysis of the geometry in Fig.~\ref{fig:pic}
(a) which represents a junction of pristine graphene and a part of
graphene which has both Kane-Mele and Rashba SOC terms.

\begin{figure}[htb] \ig[width=2.8in]{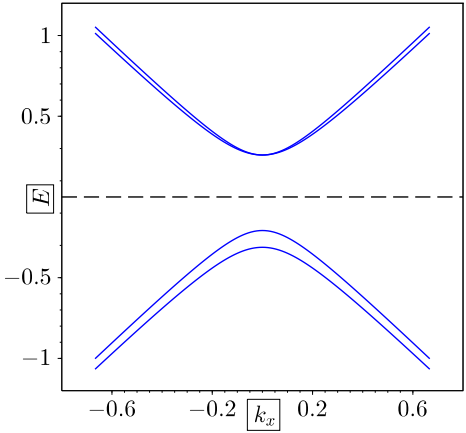}
\caption[]{Energy vs $k_x$ in a region of graphene with
Kane-Mele SOC of strength $t_2 = 0.05$, Rashba SOC of strength $\lam_R =
\De_{so}/10$, and $k_y = 0$. ($E, ~t_2$ and $\De_{so}$ are in units of $\ga$,
while $k_x$ is in units of $\ga$ and $1/d$).} \label{fig:disp} 
\end{figure}

To analyze transport across such a junction, we first consider the
system shown in Fig.~\ref{fig:pic} (a); the junction lies at $y=0$.
In region II where $y > 0$, the Hamiltonian is given by
\begin{eqnarray}
H_3 &=& \sum_{\vec k} \psi_{\vec k}^\dg h_{\vec k} \psi_{\vec k} \non \\
h_{\vec k} &=& v ~( \tau^z \si^x k_x ~+~ i \si^y
\partial_y) ~+~
\De_{so} ~\tau^z \si^z s^z \non \\
&& +~ \lam_R ~(\tau^z \si^x s^y ~-~ \si^y s^x), \label{dirham6}
\end{eqnarray}
where the momentum $k_x$ is a conserved quantity having the same
value everywhere. In the presence of the Rashba term, the
energy-momentum dispersion is given by a quartic equation for $E$,
\beq [E^2 ~-~ v^2 (k_x^2 + k_y^2) ~-~ \De_{so}^2 ]^2 ~=~ 4 \lam_R^2
~ (E - \De_{so})^2 . \label{enrel}\eeq The solution of Eq.~\eqref{enrel} 
leads to a gapped energy spectrum with four energy bands
as shown in Fig.~\ref{fig:disp} for representative values $v=3/2$,
$t_2 = 0.05$, $\lam_R = \De_{so}/10$, and $k_y = 0$. We observe that the
spectrum is not symmetric about $E=0$.

Given some values of $E$ and $k_x$ (which remain the same in the regions of
pristine graphene and the strip region with SOC), the momentum $k_y'$ in the
strip can take four values given by
\beq k_y' ~=~ \pm \frac{1}{v} ~\sqrt{E^2 - v^2 k_x^2 - \De_{so}^2 \pm 2
\lam_R |E- \De_{so}|}, \label{kyeq}\eeq
where the $\pm$ sign outside the square root is independent of the $\pm$ sign
inside. We thus have four possible values of the momentum $k_y'$. Depending
on the different parameters some of these values may be imaginary. If they
are imaginary we will consider only the exponentially decaying solutions,
while if they are real, we will choose the signs so that the group velocity
$dE/dk_y'$ is positive so that the electrons are moving right, i.e., towards
$y= \infty$. In any case, only two out of the four possible values of $k_y'$
are physically allowed in region II; let us denote these two values by
$k_y^{1,2}$.

In what follows, we further use the fact that $\tau^z$ is a good
quantum number. We will therefore only study the case $\tau^z = 1$.
The case $\tau^z = -1$ gives similar results since it is related to
$\tau^z = 1$ by the unitary transformation $h_{\vec k} \to \tau^x
\si^y h_{\vec k} \tau^x \si^y$. The operator $s^z$ is {\it not} a
good quantum number. However, we observe that $h(-k_x) = \si^y s^x
h(k_x) \si^y s^x$. Since this transformation flips both $s^z$ and
$k_x$, it is enough to study the case of an incident electron with
$s^z = 1$ and all values of $k_x$.

In region I of Fig.~\ref{fig:pic} (a), the Hamiltonian is given by
Eq.~\eqref{dirham6} with $\De_{so}=\lam_R=0$. In this region, $s^z$
is also a good quantum number. The wave functions for right and left
moving spin-up ($s^z = 1$) and spin-down ($s^z = -1$) electrons with
momentum $(k_x, \pm k_y)$ (where $k_y > 0$) and energy $E=
v\sqrt{k_x^2 + k_y^2}$ are given by \bea \psi_{\pm \ua} &=&
\frac{1}{\sqrt{2}}\left(
\begin{array}{c}
1 \\
e^{\pm i \al} \\
0 \\
0 \\
\end{array} \right) e^{i(\pm k_y y + k_x x - E t)}, \non \\
\psi_{\pm \da} &=& \frac{1}{\sqrt{2}} \left( \begin{array}{c}
0 \\
0 \\
1 \\
e^{\pm i \al} \\
\end{array} \right) e^{i(\pm k_y y + k_x x - E t)}, \label{refwav} \eea
where $e^{i\al} = (k_x - i k_y)/\sqrt{k_x^2 + k_y^2}$.
We now consider a spin-up electron which is incident on the junction
with momentum $(k_x,k_y)$ and energy $E = \mu$, where $\mu = E_F +
eV$ is the chemical potential or voltage applied in region I
measured with respect to the Dirac point and $E_F$ is the Fermi
energy. The reflected wave function can then be written as $\psi_r =
r_{\ua \ua} \psi_{- \ua} + r_{\da \ua} \psi_{-\da}$, where $r_{\ua
\ua}$ and $r_{\da \ua}$ are functions of $E$ and $k_x$. Note that
$r_{\da \ua}$ represents the amplitude for an incident spin-up
electron to be reflected from the junction as a spin-down electron.
It is therefore a direct measure of the spin active nature of the
junction. Such a reflection process which converts a spin-up
electron to a spin-down electron constitutes an analog in spin space
of Andreev reflection from a superconductor in which an incident
electron is converted to a reflected hole. The total wave function
in region I can thus be written as \bea \psi_I &=& \psi_{+\ua} ~+~
r_{\ua \ua} \psi_{- \ua} ~+~ r_{\da \ua} \psi_{- \da}.
\label{reg1wav} \eea

In region II, the presence of the Rashba term implies that $s^z$ is
not a good quantum number. Consequently, the transmitted wave
function will have amplitudes in both $s^z=1$ and $s^z = -1$
sectors. For transmitted electrons with energy $E$ and momentum
$(k_x,k_y^1)$ or $(k_x,k_y^2)$, the wave functions can be found by
solving the equation $h \psi_a = E \psi_a$, where $a=1,2$. Note that
$k_y^{1,2}$ can be real or imaginary. A straightforward calculation yields 
\bea \psi_+^a &=& \frac{1}{N_+^a} ~\left( \begin{array}{c}
u_{A \ua}^a \\
u_{B\ua}^a \\
u_{A \da}^a \\
u_{B \da}^a \\ \end{array} \right) e^{i(k_y^a y +k_x x - E t)}, \non \\
u_{A \ua}^a &=& - \frac{i \al^a}{2 \lam_R (E - \De_{so})} ~
\frac{\ep_-^a}{\ep_+^a}, \non \\
u_{B \ua}^a &=& - \frac{i \al^a}{2 \lam_R \ep_+^a}, \non \\
u_{A \da}^a &=& \frac{E-\De_{so}}{\ep_+^a}, \quad
u_{B \da}^a ~=~ 1, \label{coeffeq1} \\
\al^a &=& E^2 ~-~ v^2 (k_x^2 + k_y^2) ~-~ \De_{so}^2, \non \\
\ep_+^a &=& v(k_x-i k_y^a), \quad \ep_-^a ~=~ v(k_x+i k_y^a), \eea
where $N_+^a$ is a normalization constant which ensures that
$\psi^{\ast a}_+ \psi_+^a =1$. (The value of $N_+^a$ is not required
in the expressions presented below). The transmitted wave function
in region II is thus given by $\psi_{II} = \sum_{a=1,2} t_a
\psi_+^a$.

\begin{center} \begin{figure}[htb]
\subfigure[]{\ig[width=2.5in]{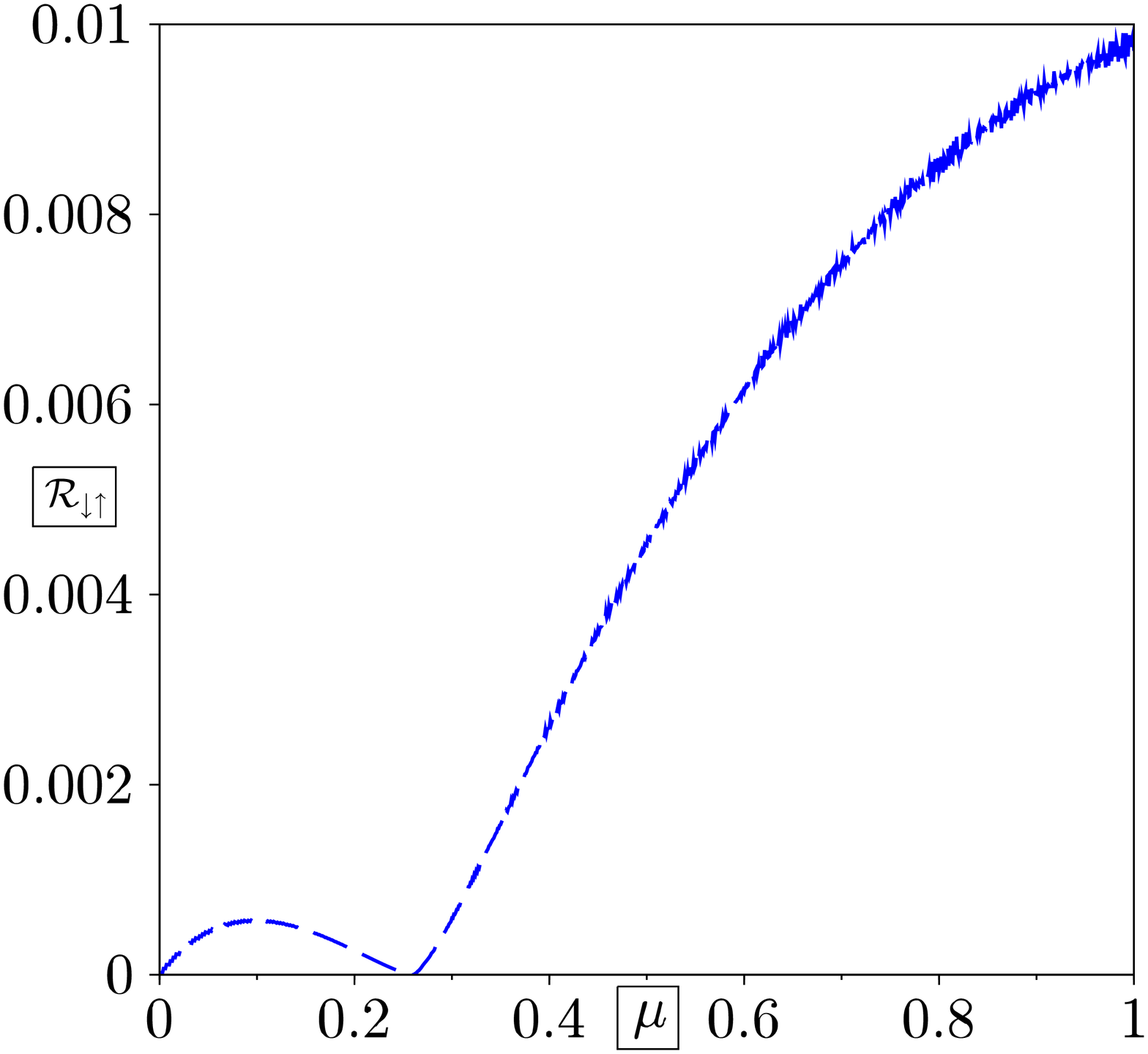}} \\ 
\subfigure[]{\ig[width=2.5in]{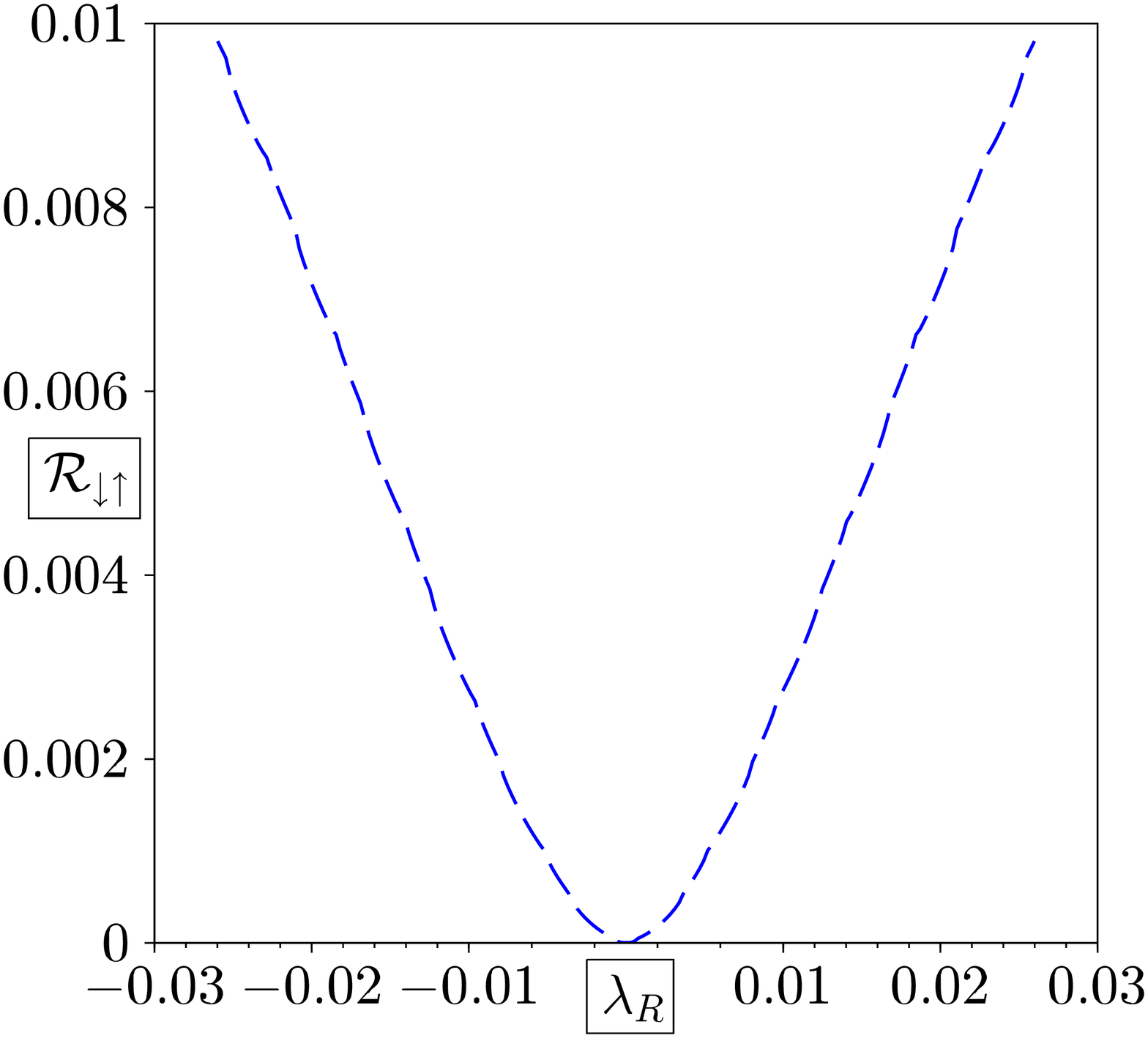}} 
\caption{Reflection probability ${\cal R}_{\da \ua}$ from a
junction between pristine graphene and graphene with a Kane-Mele SOC of 
strength $t_2 = 0.05$ and a Rashba SOC of strength $\lam_R$. Figure (a) shows
${\cal R}_{\da \ua}$ as a function of $\mu$ for $\lam_R = \De_{so}/10$, while 
figure (b) shows ${\cal R}_{\da \ua}$ as a function of $\lam_R$ for $\mu = 1$.
($\mu$ and $\lam_R$ are shown in units of $\ga$).} \label{fig19} \end{figure}
\end{center}

To find $r_{\ua \ua}$, $r_{\da \ua}$ and $t_{1,2}$, we impose continuity of
the wave function at the junction: $\psi_I(y=0)=\psi_{II}(y=0)$. This leads
to the following conditions on the various amplitudes:
\bea \frac{1+r_{\ua \ua}}{\sqrt 2} &=& \sum_{a=1,2} t_a u_{A \ua}^a,
\non \\
\frac{e^{i\al} ~+~ r_{\ua \ua} e^{-i \al}}{\sqrt 2} &=& \sum_{a=1,2}
t_a u_{B \ua}^a, \non \\
\frac{r_{\da \ua}}{\sqrt{2}} &=& \sum_{a=1,2} t_a u_{A \da}^a, \non \\
\frac{r_{\da \ua} e^{-i\al}}{\sqrt{2}} &=& \sum_{a=1,2} t_a u_{B \da}^a. \eea
The solution to these equations yields
\bea r_{\da \ua} &=& \frac{(1 - e^{i2\al}) ~(E-\De_{so}) ~
(\ep_+^2-\ep_+^1)}{\ep_+^1 \ep_+^2 ~{\mathcal D}}, \label{req2} \\
{\mathcal D} &=& e^{i\al} (u_{B \ua}^2 u_{B \da}^1- u_{B \ua}^1 u_{B \da}^2)
+ e^{-i \al}(u_{A \ua}^2 u_{A \da}^1- u_{A \ua}^1 u_{A \da}^2) \non \\
&& + (u_{A \ua}^1 u_{B \da}^2 + u_{A \da}^2 u_{B \ua}^1 -
u_{A \ua}^2 u_{B \da}^1- u_{A \da}^1 u_{B \ua}^2). \eea
Note that a non-zero value of $r_{\da \ua}$ is a consequence of the
presence of two solutions, with $k_y' = k_y^{1, 2}$, for a fixed energy $E$
and transverse momentum $k_x$; these two solutions merge when $\lam_R =0$ and
the junction ceases to be spin active in this limit. We observe that
$r_{\da \ua}=0$ for all values of $k_x$ if
$E=\De_{so}$; this shows that the specific voltage at which the spin-flip
transport takes place can be controlled by the gap originating from the
Kane-Mele term. ($r_{\da \ua}=0$ also vanishes if the incident electron comes
in at a glancing angle, namely, if $k_y = 0$ so that $e^{i\al} = \pm 1$).
For the incident electron, the range of values of $k_x$ goes from $-k_0$ to
$k_0$, where $k_0 = |\mu|/v$, since we want $E_{k_x,k_y}$ to be equal to $\mu$
with real values of $k_y$. Integrating over this range of $k_x$,
we find that the total incoming spin-up current ${\cal I}_\ua$
and the reflected spin-down current ${\cal R}_{\da \ua}$ are given by
\bea {\cal I}_\ua &=& \int_{-k_0}^{k_0} ~dk_x ~(1-|r_{\ua \ua}|^2), \non \\
{\cal R}_{\da \ua} &=& \int_{-k_0}^{k_0} ~dk_x ~ |r_{\da \ua}|^2. \eea
We show plots of ${\cal R}_{\da \ua}$ as a function of the applied voltage
$\mu$ for a fixed $\lam_R = \De_{so}/10$ in Fig.~\ref{fig19} (a), and as a
function of $\lam_R$ for a fixed $\mu=1$ in Fig.~\ref{fig19}
(b). These clearly demonstrate the spin active nature of the junction. We see
that for a fixed $\lam_R$, ${\cal R}_{\ua \da}$ indeed vanishes at $\mu =
\De_{so}$, but it eventually increases with $\mu$. This demonstrates that
the spin current can be electrically controlled.

Next, we analyze the geometry shown in Fig.~\ref{fig:pic} (b). In this
geometry, pristine graphene resides in regions I and III, and graphene with
Kane-Mele and Rashba SOC forms an interface region II of width $d$ which lies
between those two regions. To analyze the transport in this system,
we note that in regions I and III, $s^z$ is a good quantum number. In region
I, the wave function is given by Eq.~\eqref{reg1wav}. Similarly, in region
III, the transmitted wave function is given by
\bea \psi_{III} &=& t_{\ua \ua}\psi_{+ \ua} ~+~ t_{\da \ua} \psi_{+ \da},
\label{transwav3} \eea
where $t_{\si \ua}$ denotes the probability for a spin-$\si$
electron to be transmitted when the incident electron has $s^z=1$.

In region II, the electron wave function is a linear superposition of
electrons with momenta $\pm k_y^{1,2}$. The wave function of an electron
with momentum $k_y^{1,2}$ is given by Eq.~\eqref{coeffeq1},
while that for an electron with momentum $-k_y^{1,2}$ is given by
\bea \psi_-^a &=& \frac{1}{N_-^a} ~\left( \begin{array}{c}
v_{A \ua}^a \\
v_{B \ua}^a \\
v_{A \da}^a \\
v_{B \da}^a \\ \end{array} \right) e^{i(-k_y^a y +k_x x - E t)}, \non \\
v_{A \ua}^a &=& - \frac{i \al^a}{2\lam_R (E- \De_{so})} ~
\frac{\ep_+^a}{\ep_-^a}, \non \\
v_{B \ua}^a &=& - \frac{i \al^a}{2 \lam_R \ep_-^a}, \non \\
v_{A \da}^a &=& \frac{E-\De_{so}}{\ep_-^a}, \quad
v_{B \da}^a ~=~ 1, \label{coeffeq2} \eea
where $N_-^a$ is a normalization constant which ensures $\psi_-^{\dg a}
\psi_-^a=1$. Using Eqs.~\eqref{coeffeq1} and \eqref{coeffeq2}, the wave
function in region II can be written as
\bea \psi_{II} ~=~ \sum_{a=1,2} ~(p_a \psi_+^a ~+~ q_a \psi_-^a).
\label{wavreg2} \eea
The amplitudes $p_a$, $q_a$, $t_{\si \ua}$, and $r_{\si \ua}$ can be
found by matching the wave functions at $x=0$ and $x=d$. This yields
\bea \frac{1+r_{\ua \ua}}{\sqrt 2} &=& \sum_{a=1,2} (p_a u_{A \ua}^a + q_a
v_{A \ua}^a), \non \\
\frac{e^{i \al} ~+~ r_{\ua \ua} e^{-i \al}}{\sqrt 2} &=& \sum_{a=1,2}
(p_a u_{B \ua}^a + q_a v_{B \ua}^a), \non \\
\frac{r_{\da \ua}}{\sqrt 2} &=& \sum_{a=1,2} (p_a u_{A \da}^a + q_a
v_{A \da}^a), \non \\
\frac{r_{\da \ua} e^{-i \al}}{\sqrt 2} &=& \sum_{a=1,2} (p_a u_{B
\da}^a + q_a v_{B \da}^a), \label{bcond2} \\
\frac{t_{\ua \ua} e^{i k_y d}}{\sqrt 2} &=& \sum_{a=1,2} (p_a u_{A \ua}^a
e^{i k_y^a d} + q_a v_{A \ua}^a e^{-i k_y^a d}), \non \\
\frac{t_{\ua \ua} e^{i (k_y d + \al)}}{\sqrt 2} &=& \sum_{a=1,2} (p_a
u_{B \ua}^a e^{i k_y^a d} + q_a v_{B \ua}^a e^{-i k_y^a d}), \non \\
\frac{t_{\da \ua} e^{i k_y d}}{\sqrt 2} &=& \sum_{a=1,2} (p_a u_{A \da}^a
e^{i k_y^a d} + q_a v_{A \da}^a e^{-i k_y^a d}), \non \\
\frac{t_{\da \ua} e^{i (k_y d + \al)}}{\sqrt 2} &=& \sum_{a=1,2} (p_a
u_{B \da}^a e^{i k_y^a d} + q_a v_{B \da}^a e^{-i k_y^a d}). \non \\
\label{bcond3} \eea
{} Using these we can compute the reflection and transmission probabilities
$R_{\al \ua} = |r_{\al \ua}|^2$ and $T_{\al \ua} = |t_{\al \ua}|^2$
respectively, where $\al$ can be $\ua$ or $\da$. Similarly, if the electron
incident from region I had $s^z =-1$, we would have the reflection and
transmission probabilities $R_{\al \da}$ and $T_{\al \da}$.
These must satisfy the unitarity relations
\bea T_{\ua \ua} ~+~ T_{\da \ua} ~+~ R_{\ua \ua} ~+~R_{\da \ua} ~=~ 1, \non \\
T_{\ua \da} ~+~ T_{\da \da} ~+~ R_{\ua \da} ~+~R_{\da \da} ~=~ 1.
\label{unit} \eea

In what follows, we shall compute the tunneling conductance by
solving Eqs.~(\ref{bcond2}-\ref{bcond3}). and compute the transmission
probabilities between two pristine graphene regions (region I and
III) across a strip of graphene (region II) with a width $d_0= 15d$,
for $t_2 = 0.05$ and $\lam_R = \De_{so}/10$. In Figs.~\ref{fig21}
(a) and (b), we show the transmission probabilities $T_{\ua \ua} =
|t_{\ua \ua}|^2$ and $T_{\da \ua} = |t_{\da \ua}|^2$ versus $k_x$
for two values of the energy $E$. Our plots clearly demonstrates a
finite spin conversion as indicated by the dashed blue lines in
Figs.\ \ref{fig21}(a) and (b). Given the transmission probabilities,
the differential conductances can be calculated as follows. For a
momentum $\vk = (k_x, k_y)$, the current in the $y$ direction is
given by $J_{y,\vk} ~=~ |dE_{\vk}/ dk_y|$. Let the chemical
potentials in regions I and III in Fig.~\ref{fig:pic} (b) be $\mu_1$
and $\mu_2$, so that the voltage bias between the two regions is
given by $eV= (\mu_2 - \mu_1)/e$. In the zero bias limit in which $V
\to 0$ and $\mu_1, ~\mu_2 \to \mu$, the differential conductance
$G_{\al \be} =dI/dV$ for an incident electron with spin $\al$ being
transmitted with spin $\be$ ($\al, \be$ can be $\ua$ or $\da$) is
given by \beq G_{\al \be} (\mu) ~=~ e^2 W ~\int \int ~\frac{dk_x
dk_y}{(2\pi)^2} ~\de (\mu - E_{\vk}) ~T_{\al \be} ~J_{y,\vk},
\label{diff1} \eeq where $W$ is the width of the system in the $x$
direction (we assume that $W \gg d$). Integrating the $\de$-function
over $k_y$ in Eq.~\eqref{diff1} gives a denominator equal to
$|(dE_{\vk}/ dk_y)_{E_{\vk} = \mu}|$ which precisely cancels the
$J_{y,\vk}$ appearing in the numerator of that equation. We thus
obtain \beq G_{\al \be} (\mu) ~=~ \frac{e^2 W}{(2\pi)^2}
~\int_{-k_0}^{k_0} ~dk_x ~ T_{\al \be}, \label{diff2} \eeq where
$k_0 = |\mu|/v$.

Instead of plotting $G_{\al \be} (\mu)$ versus $\mu$, it is convenient to
plot the ratio $G_{\al \be} (\mu) /G_0 (\mu)$, where $G_0 (\mu)$ is the
conductance when there is perfect transmission, i.e., $T_{\al \be} =
\de_{\al \be}$. From Eq.~\eqref{diff2} we find that $G_0 (\mu) = [e^2
W/(2\pi)^2] [2|\mu|/v]$ in one particular valley. We then have the expressions
\bea \frac{G_{\ua \ua}}{G_0} &=& \frac{1}{2k_0} ~\int_{-k_0}^{k_0} ~dk_x ~
T_{\ua \ua}, \non \\
\frac{G_{\da \ua}}{G_0} &=& \frac{1}{2k_0} ~\int_{-k_0}^{k_0} ~dk_x ~
T_{\da \ua}. \eea
Fig.~\ref{fig21} (c) shows plots of $G_{\ua \ua}/G_0$ and $G_{\da \ua}/G_0$
versus $\mu$.

\begin{widetext} \begin{center} \begin{figure}[htb]
\subfigure[]{\ig[width=2.3in]{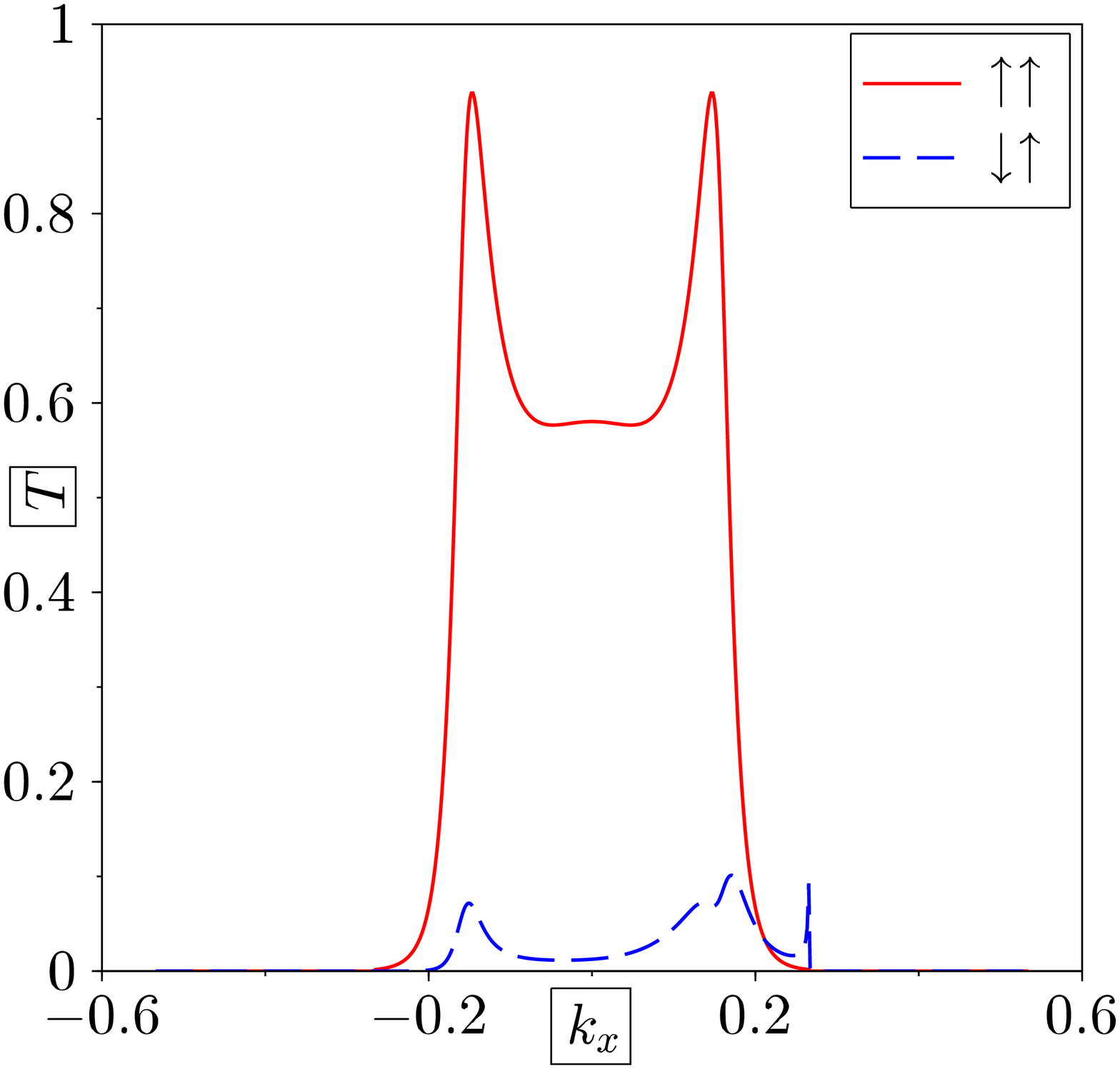}} 
\subfigure[]{\ig[width=2.3in]{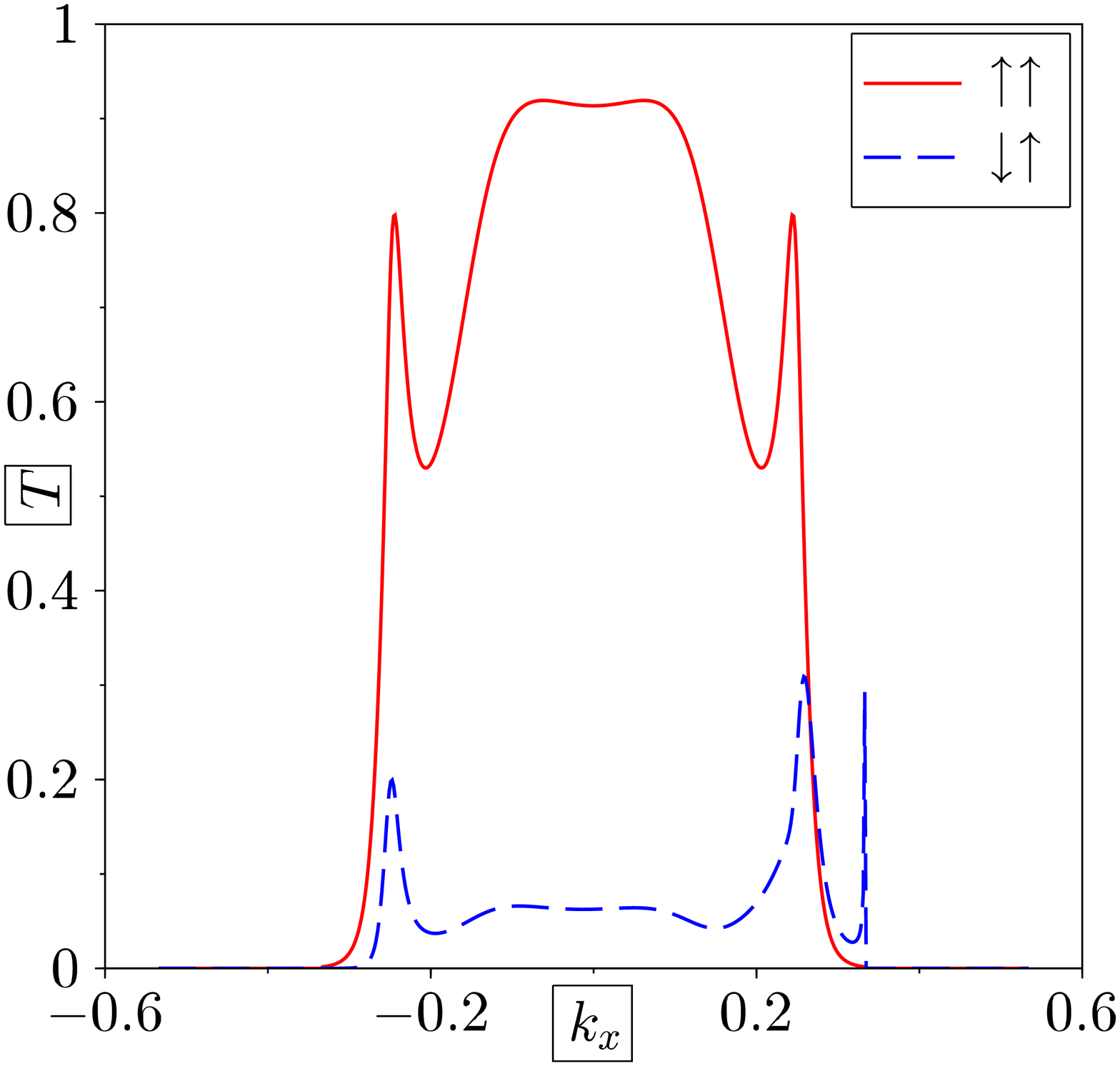}} 
\subfigure[]{\ig[width=2.3in]{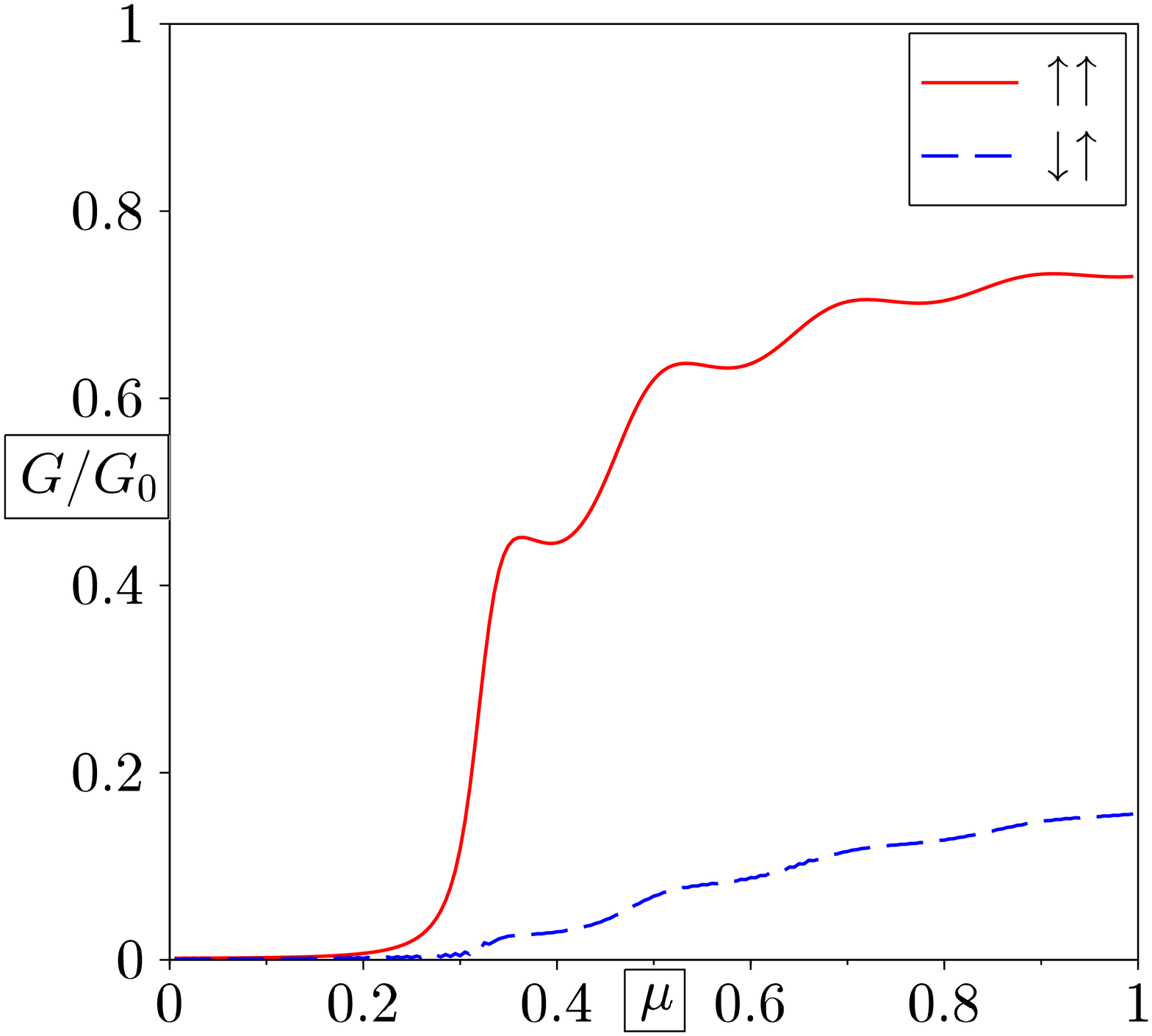}} 
\caption{Transmission probabilities and conductance across a
strip of graphene with a width of 15 (in units of the lattice spacing $d$), a
Kane-Mele SOC of strength $t_2 = 0.05$, and a Rashba SOC of strength $\lam_R
= \De_{so}/10$. (a) and (b) show the transmission probabilities $T_{\ua \ua}$
(solid red) and $T_{\da \ua}$ (dashed blue) vs $k_x$ for $E=0.4$ and $0.5$
respectively, while (c) shows $G_{\ua \ua}/G_0$ (solid red) and $G_{\da \ua}$
(dashed blue) vs $\mu$. ($t_2$ and $\De_{so}$ are in units of $\ga$, 
while $k_x$ is in units of $1/d$).} \label{fig21} \end{figure} \end{center} 
\end{widetext}

\begin{figure}[htb] \ig[width=2.8in]{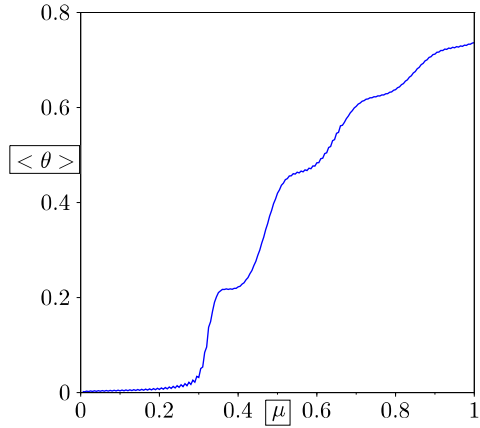} 
\caption[]{Average rotation angle vs $\mu$ for transmission
across a strip of graphene with a width of 15 (in units of the lattice spacing
$d$), a Kane-Mele SOC of strength $t_2 = 0.05$, and a Rashba SOC of strength
$\lam_R = \De_{so}/10$. ($\mu, ~t_2$ and $\De_{so}$ are in units of 
$\ga$).} \label{fig:ang} \end{figure}

Another interesting quantity to consider is the rotation of the
electron spin produced by region II. For each value of $E$ and
$k_x$, we know that a spin-up electron incident from region I
converts to a linear superposition of spin-up and spin-down on being
transmitted to region III, with amplitudes $t_{\ua \ua}$ and $t_{\da
\ua}$ respectively. In spin space, the linear superposition $(t_{\ua
\ua}, t_{\da \ua})^T$ describes an electron whose spin polarization
points at an angle $\ta$ with respect to the $z$ axis, where $\tan
(\ta/2) = |t_{\da \ua}/t_{\ua \ua}|$. We can therefore define an
average rotation angle produced by region II as
\beq \la \ta \ra ~=~ \frac{1}{2k_0} ~\int_{-k_0}^{k_0} ~dk_x ~\ta ~(
T_{\ua \ua} ~+~ T_{\da \ua}), \eeq
where we have weighted the angle of rotation by
the transmission probability $T_{\ua \ua} + T_{\da \ua}$. In
Fig.~\ref{fig:ang} we show the average rotation angle as a function
of $\mu$ for transmission across a strip of graphene with the same
parameters as in Fig.~\ref{fig21}. Fig.~\ref{fig:ang} clearly shows a
finite spin rotation which increases as a function of $\mu$ in the
zero-bias limit; this demonstrates the potential of these junctions
as generators of electrically controllable spin current.

Before ending this section, we note that the calculations of this
section serve as a proof of principle that a finite electrically
controllable spin current may be generated in graphene junctions
with finite SO coupling. There are many ways to enhance the
magnitude of this current, which in our chosen parameter regime,
appears to be rather small. For example, one can increase the bias
voltage $V$ and/or the thickness $d$ of region II which will
increase both $T_{\uparrow \downarrow}$ and $\theta$. Further, there
may be other, more suitable geometries for larger spin current
generation. In the next sub-section, we shall carry out numerical
calculation from a lattice Hamiltonian which will address some of
these issues and verify the approximate continuum calculation of the
present section.

\subsection{Numerical calculation using lattice models}
\label{sec:condnum}

In this section, we provide a numerical analysis of the transport across
the various junctions starting from a lattice model. The advantages of a 
lattice calculation over a continuum calculation are that a lattice calculation
is straightforward to implement numerically (for instance, one does not have 
to impose any matching conditions on the wave functions), and one can easily
study the effects of potentials or spin-orbit couplings which have arbitrary 
spatial profiles.

We will follow a procedure similar to the study of conductance across 
junctions of graphene and other materials~\cite{sengupta06,bhattacharjee06}. 
We will assume that the strip is infinitely long in the $x$ direction,
so that the momentum $k_x$ of an electron incident from one of the
regions of pristine graphene is a good quantum number everywhere in
the system. The incident energy $E$ is also a good quantum number.
However, $k_y$ will vary from one region to another depending on the
presence of SOC and a Zeeman field.

We will calculate the conductance numerically using a lattice model
similar to the one shown in Fig.~\ref{fig03}. We consider an
electron incident from the pristine graphene at the bottom of that
figure and we calculate the probabilities of reflection (back to the
bottom) and transmission (to the pristine graphene at the top). An
incident spin-up electron can get either transmitted
or reflected as spin-up or spin-down; we will denote the corresponding
probabilities by $T_{\ua \ua}$, $T_{\da \ua}$, $R_{\ua \ua}$, and
$R_{\da \ua}$, respectively. Similarly an incident spin-down electron
will have transmission and reflection probabilities given
by $T_{\ua \da}$, $T_{\da \da}$, $R_{\ua \da}$, and $R_{\da \da}$.

The calculation is done as follows. Given the values of the momentum $k_x$
and energy $E$ (which we will henceforth assume to be positive), the
dispersion for pristine graphene given in Eq.~\eqref{ek} uniquely fixes a
momentum $k_y$ lying in the range $[0,2\pi/3d]$. (It may happen that there is
no real solution for $k_y$; this would imply that such a value of $E$ is not
allowed for the given momentum $k_x$. In that case we will set the 
transmission probabilities equal to zero). Then the incident and transmitted
waves will have momentum $k_y$ while the reflected wave will have
momentum $-k_y$. We now consider a single transmitted wave, with
unit amplitude and $s^z$ equal to either $1$ or $-1$, which is
located at the top of Fig.~\ref{fig03}, and we find which
superposition of the four possible incident and reflected waves at
the bottom would give rise to such a transmitted wave (we have to
allow for four possible waves in general since they could be either
incident or reflected and they could have $s^z = \pm 1$). This
superposition can be found by using Eqs.~\eqref{eom} to set up a
matrix problem where the four reflection and incident amplitudes as
well as the values of $a_m$ and $b_m$ inside the region with SOC or
Zeeman field appear on the left side of an equation and the single
transmitted wave at the top (with unit amplitude) appears as a
source term on the right of the equation; the reflection and
incident amplitudes are then found by doing a matrix inversion.
Having found these amplitudes for the two cases where the
transmitted wave has $s^z$ equal to $1$ and $-1$, we then invert
these relations and find the reflection and transmission amplitudes
when a wave is incident with unit amplitude $s^z = \pm 1$. The
modulus squared of the amplitudes give the reflection and
transmitted probabilities as usual. Finally we check if the
unitarity relations in Eq.~\eqref{unit} are satisfied.

Given the transmission probabilities, the differential conductances can
be calculated as described in Sec.~\ref{sec:condan}. We again arrive at
Eqs.~\eqref{diff1} and \eqref{diff2}, except that the range of integration of
$k_x$ in the lattice model is given by $[-2\pi/\sqrt{3},2\pi/\sqrt{3}]$.
However only those values of $k_x$ will contribute for which $E_{\vk}$
can be equal to $\mu$ with real values of $k_y$. Once again, we will
plot the ratio $G_{\al \be} (\mu) /G_0 (\mu)$, where $G_0 (\mu)$ is the
conductance when $T_{\al \be} = \de_{\al \be}$. Given a chemical potential
$\mu$ lying between 0 and 1, we can show using Eq.~\eqref{ek} that
\bea G_0 (\mu) &=& \frac{e^2 W}{(2\pi)^2} ~\int ~dk_x \non \\
&=& \frac{e^2 W}{(2\pi)^2} ~\frac{8}{\sqrt{3}} ~[ {\rm acos} (\frac{1
- \mu}{2}) ~-~ {\rm acos} (\frac{1 + \mu}{2}) ]. \non \\
&& \label{diff3} \eea
If $\mu$ is small, $G_0 (\mu)$ varies linearly
with $\mu$, namely, $G_0 (\mu) = [e^2 W/(2\pi)^2] [8\mu/v]$ where
$v=3/2$ is the Fermi velocity. This expression is exactly twice of
what we expect for two species (due to the valleys) of massless
Dirac electrons in the continuum. The additional factor of two is
because we have considered the full range of $k_x$ from
$[-2\pi/\sqrt{3}$ to $2\pi/\sqrt{3}]$; this double counts the
contribution from each of the two valleys since the transmission is
invariant under $k_x \to k_x + 2\pi/ \sqrt{3}$. The double counting
is not present in the ratio $G_{\al \be} (\mu)/ G_0 (\mu)$.

\begin{widetext} \begin{center} \begin{figure}[htb]
\subfigure[]{\ig[width=2.3in]{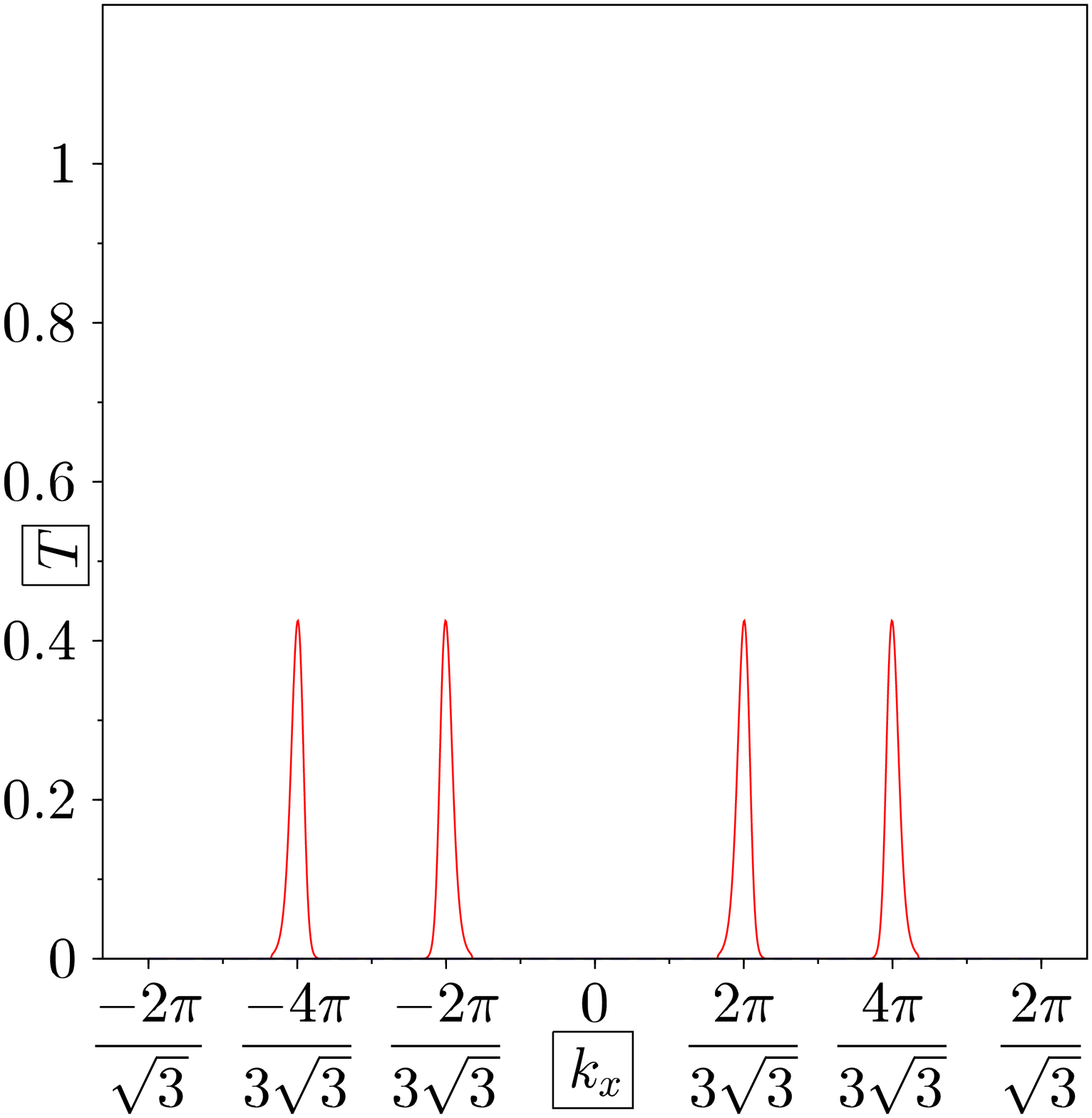}}
\subfigure[]{\ig[width=2.3in]{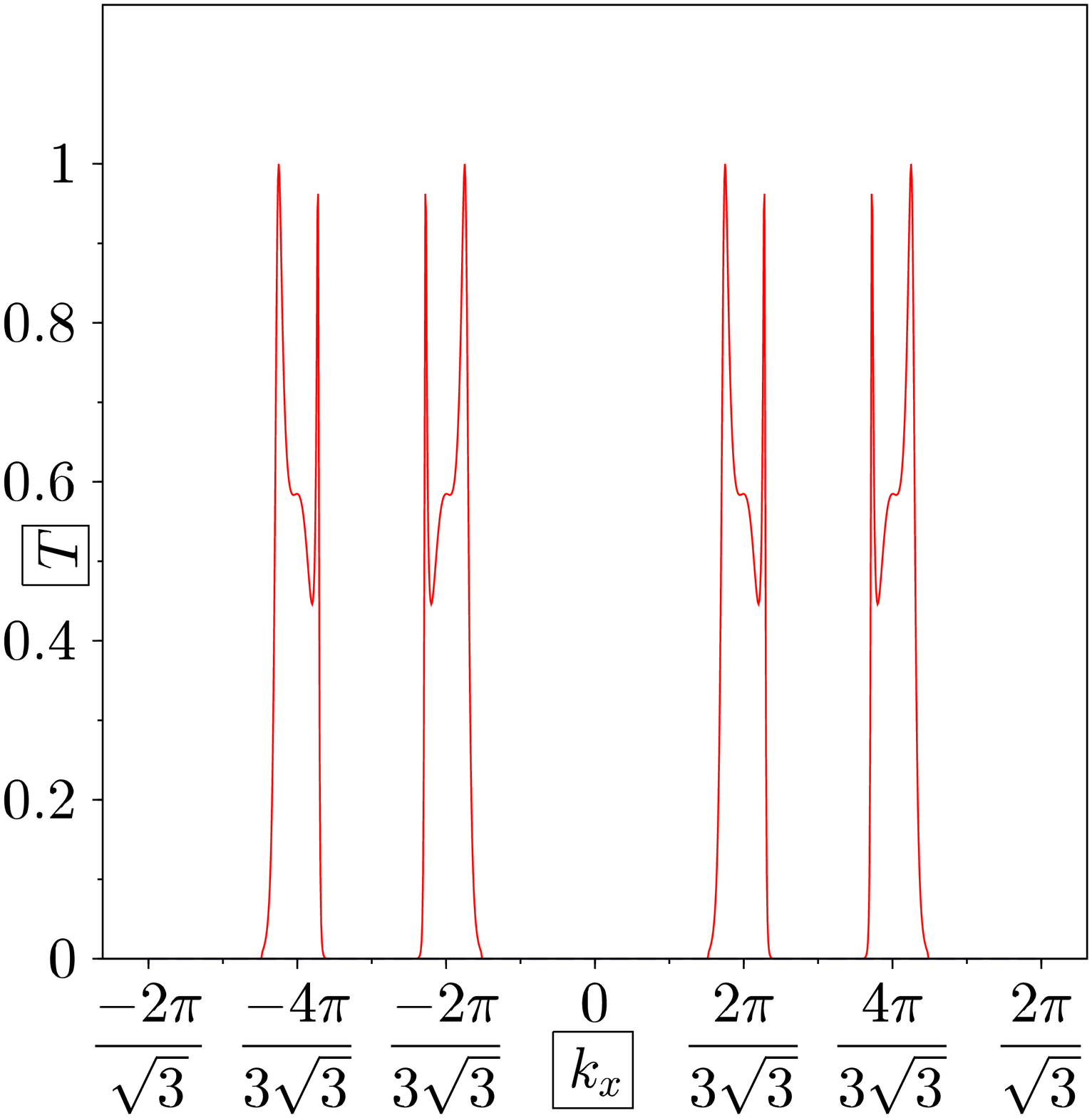}}
\subfigure[]{\ig[width=2.3in]{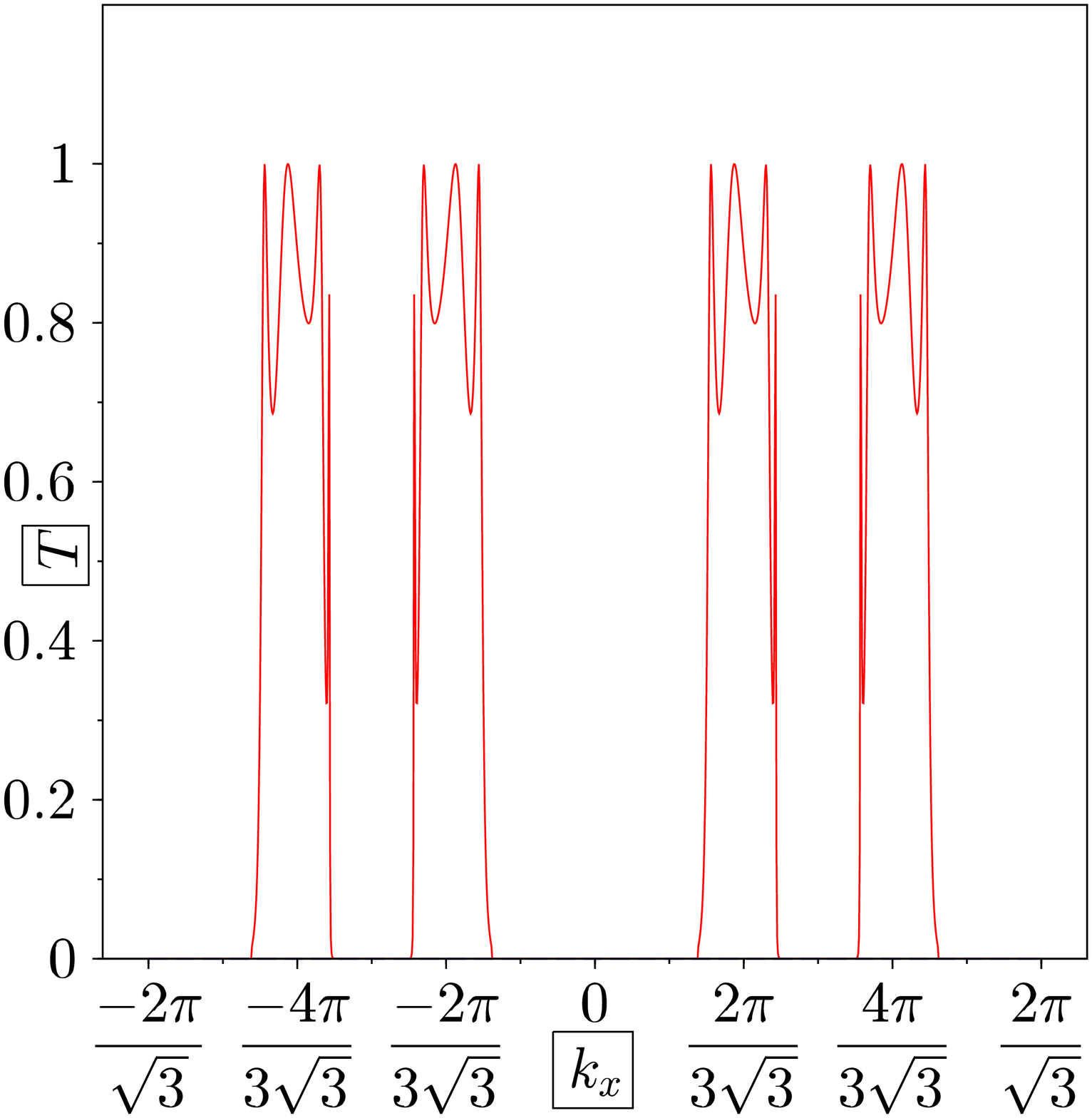}} 
\caption{Transmission probability $T_{\ua \ua}$ vs $k_x$ 
across a strip of graphene with 15 unit cells and a SOC of strength $t_2 = 
0.05$. The values of energy are (a) $E=0.3$, (b) $E=0.4$, and (c) $E=0.5$. 
($E$ and $t_2$ are in units of $\ga$, while $k_x$ is in units of $1/d$).} 
\label{fig07} \end{figure} \end{center} \end{widetext}

\begin{figure}[htb] \ig[width=3.1in]{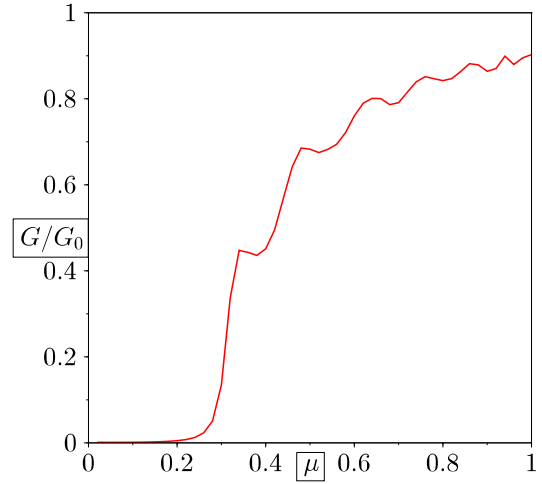}
\caption[]{$G_{\ua \ua} /G_0$ vs $\mu$ for transmission 
across a strip of graphene with 15 unit cells and a SOC of strength $t_2 = 
0.05$. ($\mu$ and $t_2$ are in units of $\ga$).} \label{fig71} 
\end{figure}

We now present our numerical results for a number of different
cases. We first consider a strip of graphene with a width of 15 unit
cells and a SOC of strength $t_2 = 0.05$; there is pristine graphene
on both sides of the strip. (We take the Zeeman field to be zero).
We will study the transmission probabilities $T_{\al \be}$ as a
function of the incident energy $E$ and the momentum $k_x$. Since
the SOC does not couple spin-up and spin-down electrons, we will have
$T_{\da \ua} = T_{\ua \da} = 0$. Further, using the symmetries
discussed after Eqs.~\eqref{eom}, we can show that
\bea T_{\ua \ua} (k_x) &=& T_{\ua \ua} (k_x + 2 \pi/\sqrt{3}), \non \\
T_{\ua \ua} (k_x) &=& T_{\da \da} (-k_x). \label{sym1} \eea
The second equation in \eqref{sym1} implies that it is sufficient to
study $T_{\ua \ua}$. In Fig.~\ref{fig07}, we show the transmission
probability $T_{\ua \ua}$ as a function of $k_x$ for three values of
the energy, $E= 0.3, ~0.4$ and $0.5$. In each of the figures we see
that there are regions of $k_x$ where $T_{\ua \ua}$ is exactly zero
or close to zero. These regions occur for two reasons. First, we
have already seen that in pristine graphene, for a given value of
$k_x$, all possible values of $E$ are not allowed; for a disallowed
value of $E$, we set $T_{\ua \ua} = 0$. Second, in graphene with a
SOC of strength $t_2 = 0.05$, the minimum value of energy occurs at
the four values $k_x = \pm 2 \pi/(3 \sqrt{3})$ and $\pm 4 \pi/(3
\sqrt{3})$ and that minimum energy is given by $\De_{so} = 3
\sqrt{3} t_2 \simeq 0.26$. In Fig.~\ref{fig07} (a), the energy $E =
0.3$ is only a little bit more than $\De_{so}$. Hence for all values
of $k_x$ except the regions around the four special momenta, the
energy of the incident electron lies inside the gap of graphene with
SOC, and the wave function will decay exponentially inside that part
of graphene. $T_{\ua \ua}$ is therefore very small for all values of
$k_x$ except near those four momenta.

In Fig.~\ref{fig07} we see some transmission resonances, namely, for
certain values of $E$ and $k_x$, we find that $T_{\ua \ua}$ is close
to 1. We can understand this as follows. If the energy does not lie
in the gap of region of graphene with SOC, i.e., if the momentum
$k_y' = \pm (1/v) \sqrt{E^2 - v^2 k_x^2 - \De_{so}^2}$ in that
region is real, then we expect a transmission resonance if $(3/2)
k_y' N_y$ is an integer multiple of $\pi$. This is because such a
condition implies that the wave function in the region with SOC will
satisfy $\psi (y=N_y) = \pm \psi (y=0)$, where the $\pm$ sign
depends on whether $(3/2) k_y' N_y$ is an even or odd multiple of
$\pi$. Hence the wave function will match at $y=0$ and $N_y$ between
pristine graphene and graphene with SOC, with the reflection
amplitude being equal to zero at $y=0$ and the transmission
amplitude being equal to $\pm 1$ at $y=N_y$; we will therefore get
$T_{\ua \ua} = 1$.

In Fig.~\ref{fig71}, we show $G_{\ua \ua} /G_0$ as a function of
$\mu$. We see that $G_{\ua \ua} /G_0$ is very small for $\mu
\lesssim \De_{so} \simeq 0.26$. As $\mu$ is increased to 1, $G_{\ua
\ua} /G_0$ also approaches 1 although some oscillations are visible.
The locations of the maxima can be qualitatively understood as
follows. We saw in the previous paragraph that there are
transmission resonances if $(3/2) k_y' N_y = n \pi$, where
$n=1,2,3,\cdots$. Since Fig.~\ref{fig07} shows that the resonances
are most prominent close to the Dirac points $k_x = \pm 2\pi/
(3\sqrt{3})$ and $\pm 4\pi/(3\sqrt{3})$, let us ignore the
contributions from values of $k_x$ away from the Dirac points and
approximate the dispersion inside graphene with SOC by $E= \sqrt{(v
k_y')^2 + \De_{so}^2}$; this holds if $E$ is not too large. We
therefore expect $G_{\ua \ua} /G_0$ to show maxima when \beq \mu =
\sqrt{\left( \frac{2\pi n v}{3N_y} \right)^2 + \De_{so}^2}. \eeq
The smallest values of $n=1,2,3$ give $\mu=0.334,0.493,0.680$ which
are approximately the locations of the first three maxima in
Fig.~\ref{fig71}.

The system discussed above, with only a SOC present, enjoys an additional
symmetry, namely,
\bea T_{\ua \ua} (k_x) &=& T_{\da \da} (k_x). \label{sym2} \eea
In fact, even the transmission amplitudes
are equal, $t_{\ua \ua} (k_x) = t_{\da \da} (k_x)$. This can be
shown as follows. We first note that for a particular value of $s^z$
equal to either $1$ or $-1$, Eqs.~\eqref{eom} have a symmetry
resembling time reversal in which all numbers are complex
conjugated. (This does not change the value of $k_x$ which simply
appears as a parameter in those equations. This symmetry is
therefore a bit different from the usual time reversal symmetry in
which both $k_x$ and $k_y$ change sign). For a particular value of
$s^z$ equal to $1$ or $-1$, this implies that the scattering matrix
$S$ which relates the incoming waves at the top and bottom of the
system to the outgoing waves must be symmetric, in addition to being
unitary. (This can be proved as follows. If $i_1$ and $i_2$ denote
the incoming amplitudes at the top and bottom, with plane wave
factors $e^{-i 3k_y/2}$ and $e^{i 3k_y/2}$, and $o_1$ and $o_2$
denote the outgoing amplitudes at the top and bottom, with plane
wave factors $e^{i 3k_y/2}$ and $e^{-i 3k_y/2}$, they must be
related as $(o_1, o_2)^T = S (i_1, i_2)^T$. Complex conjugating this
relation transforms $i^*_{1/2}$ to $o_{1/2}$ and vice versa. Time
reversal symmetry then implies that we must have $(i_1^*, i_2^*)^T =
S (o_1^*, o_2^*)^T$. This implies that $S^\dag = S^*$, namely, $S$
is symmetric. Hence the transmission amplitude $S_{21}$ from the top
to the bottom must be equal to the transmission amplitude $S_{12}$
from the bottom to the top. Next, we use the fact that the Hamiltonian
in Eq.~\eqref{dirham3} (but without a magnetic field ${\vec b}$)
is symmetric under the parity transformation $y \to - y$, $x \to x$, namely,
\beq h_{k_x,-k_y} ~=~ \si^x ~s^x ~h_{k_x,k_y} ~\si^x ~s^x. \label{parity} \eeq
The transformation in Eq.~\eqref{parity} interchanges the $a$ and $b$
sublattices and also flips the $s^z$ component of the spin. This symmetry
implies that the transmission amplitude from the top to the bottom for an
electron with spin $s^z$ must be equal to the transmission amplitude from
the bottom to the top for an electron with spin $- s^z$. Combining these two
symmetries, we see that the transmission from the bottom to the top must be
the same for $s^z = \pm 1$.

\begin{figure}[htb] \ig[width=3.1in]{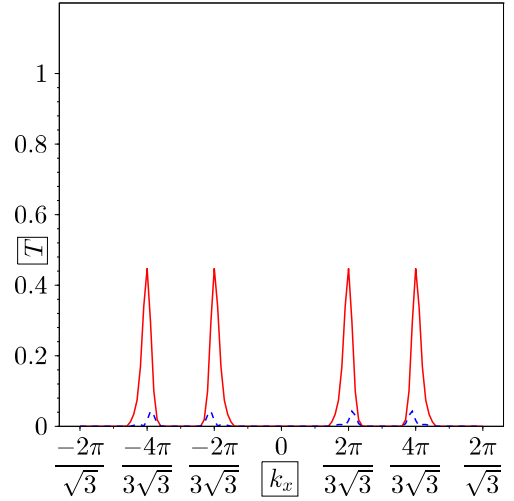}
\caption[]{$T_{\rightarrow \rightarrow}$ (solid red) and 
$T_{\leftarrow \rightarrow}/G_0$ (dashed blue) for transmission of an electron
with energy $E=0.5$ and $s^x = 1$ across a strip of graphene with 15 unit 
cells and a SOC of strength $t_2 = 0.05$; the bottom seven unit cells have $V=
-0.2$ and the top eight unit cells have $V=0.2$. ($E, ~t_2$ and $V$ are in 
units of $\ga$, while $k_x$ is in units of $1/d$).} \label{fig82} \end{figure}

\begin{figure}[htb] \ig[width=3.1in]{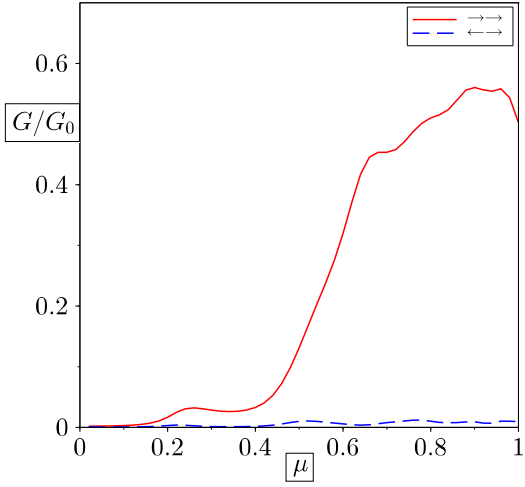}
\caption[]{$G_{\rightarrow \rightarrow} /G_0$ (solid
red) and $G_{\leftarrow \rightarrow}/G_0$ (dashed blue) vs $\mu$
for transmission across a strip of graphene with 15 unit cells and a
SOC of strength $t_2 = 0.05$; the bottom seven unit cells have $V=-0.2$ and 
the top eight unit cells have $V=0.2$. ($\mu, ~t_2$ and $V$ are in units of 
$\ga$, while $k_x$ is in units of $1/d$).} \label{fig83} \end{figure}

If the parity symmetry is broken, by applying a potential $V_m$
which depends on the $y-$coordinate $m$ in a way which is not
invariant under reflection about the center of the region with SOC
($V_m$ can be independent of both the spin and the sublattice
index), we expect that the transmission amplitudes $t_{\ua \ua}$ and
$t_{\da \da}$ will no longer be equal. Taking linear combinations of
the incident electron so as to be quantized along, say, the $x$
direction, we find that the transmission amplitudes for a $s^x = 1$
electron to be transmitted as a $s^x = 1$ and $-1$ electron are
given by $(t_{\ua \ua} + t_{\da \da})/2$ and $(t_{\ua \ua} - t_{\da
\da})/2$ respectively. (In fact these expressions hold for any
component of the spin which is perpendicular to the $z$ axis, not
just $s^x$). The latter will not be zero in general which implies
that the $s^x$ component of the electron can flip when it transmits
across a region with SOC and a parity-breaking potential. We
demonstrate this effect in Figs.~\ref{fig82} and \ref{fig83} for
transmission across a strip of graphene with a width of 15 unit
cells with a SOC of strength $t_2 = 0.05$; in addition, the bottom seven
unit cells have a potential $V = -0.2$ and the top eight unit cells have
$V=0.2$. Fig.~\ref{fig82} shows the probabilities for an incident
electron with energy $E=0.5$ and $s^x = 1$ to be transmitted into
an electron with $s^x = \pm 1$; the transmission probabilities
$T_{\rightarrow \rightarrow}$ and $T_{\leftarrow \rightarrow}$ are
shown as functions of $k_x$. (Here $\rightarrow, \leftarrow$ denote
$s^x = \pm 1$). We see that there is a non-zero (though small)
probability of conversion from $s^x = 1$ to $-1$.
Figure \ref{fig83} shows plots of $G_{\rightarrow \rightarrow}$ and
$G_{\leftarrow \rightarrow}$ versus $\mu$ for the same system. (The spin
conversion effect discussed here is related to spin filter and spin valve
effects which have been discussed in other papers, for instance,
Refs.~\onlinecite{haugen08,yang13,song15}).


\begin{figure}[htb] \ig[width=3.1in]{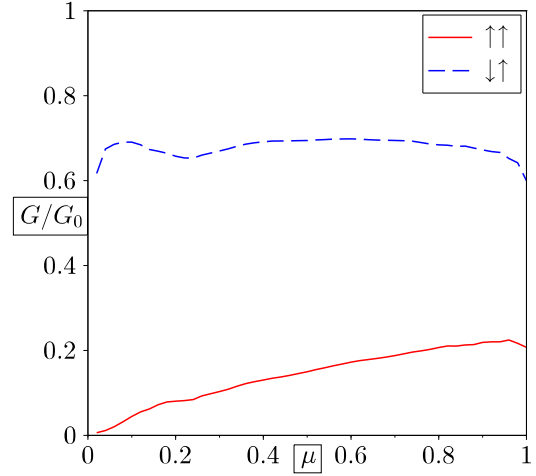}
\caption[]{$G_{\ua \ua} /G_0$ (solid red) and $G_{\da \ua}/
G_0$ (dashed blue) vs $\mu$ for transmission across a strip of graphene 
with 15 unit cells and a Zeeman field $b_x = 0.1$. ($\mu$ and $b_x$ are 
in units of $\ga$).} \label{fig81} \end{figure}

Next we consider a strip of graphene with a width of 15 unit cells and a
Zeeman field in the $x$ direction with $b_x = 0.1$. (We take the SOC
strength to be zero). To study this problem, we generalize Eqs.~\eqref{eom}
to include a Zeeman field $b_x$ which couples spins $\ua$ and $\da$; now
$T_{\ua \ua}$, $T_{\da \ua}$, $T_{\da \da}$, and $T_{\ua \da}$ will all
be non-zero in general. We can then derive some symmetries similar to the
ones discussed after Eqs.~\eqref{eom}; using these we find that
\bea T_{\al \be} (k_x) &=& T_{\al \be} (k_x + 2 \pi/\sqrt{3}), \non \\
T_{\al \be} (k_x) &=& T_{-\al, -\be} (-k_x), \label{sym3} \eea
where we define $-\al = \da (\ua)$ if $\al = \ua (\da)$ and similarly for
$-\be$ in the second equation. It is therefore enough to study $T_{\ua \ua}$
and $T_{\da \ua}$. (If $t_2 = 0$, we also have the symmetry $T_{\al \be}
(k_x) = T_{\al \be} (-k_x)$). 

In Fig.~\ref{fig81}, we show $G_{\ua \ua} /G_0$ and $G_{\da \ua}
/G_0$ as functions of $\mu$. We see that $G_{\da \ua} /G_0$ is much
larger than $G_{\ua \ua} /G_0$ in the entire range of $\mu$. This happens
for this particular value of $N_y = 15$ and can be qualitatively understood
as follows. If electrons with $s^x = \pm 1$ (rather than $s^z = \pm 1$) were
incident, they would be transmitted with unit magnitude but with a
phase difference. If the electrons have energy $E$, the dispersion
inside graphene with a Zeeman field would be given by $E= (v k_{y
\pm}) \pm b_x$, if $E$ is not too large. (For simplicity, we are assuming
that $k_x$ is equal to one of the Dirac points $\pm 2\pi/ (3\sqrt{3})$ or
$\pm 4\pi/(3\sqrt{3})$). Here $k_{y \pm}$ denote the
values of the $y$ component of the momentum for $s^x = \pm 1$. The
phase difference between the electrons with $s^x = \pm 1$ is given
by $(3/2) (k_{y-} - k_{y+}) N_y = 3 b_x N_y/v$. For $b_x = 0.1$ and
$N_y = 15$, the phase difference is $3$ which is close to $\pi$.
Hence electrons with $s^x = \pm 1$ are perfectly transmitted but
with almost opposite signs. Hence incident electrons with $s^z =
1$, which is given by the linear combination $(|s^x = 1 \ra + |s^x
= -1 \ra)/\sqrt{2}$ will be transmitted almost as the linear
combination $(|s^x = 1 \ra - |s^x = -1 \ra)/\sqrt{2}$ which is the
same as $s^z = -1$. We thus see an almost perfect conversion of spin
from $s^z = 1$ to $-1$. Note that this approximate argument is
independent of the energy $E$ which explains why $G_{\da \ua} /G_0$
is much larger than $G_{\ua \ua} /G_0$ for all $\mu$ in
Fig.~\ref{fig81}.


\begin{figure}[htb] \ig[width=3.1in]{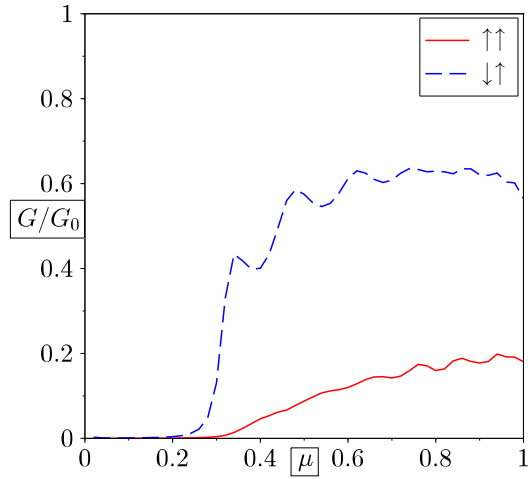}
\caption[]{$G_{\ua \ua} /G_0$ (solid red) and $G_{\da \ua}/G_0$
(dashed blue) vs $\mu$ for transmission across two successive strips of 
graphene, the first one with 15 unit cells and a SOC of strength $t_2 = 0.05$,
and the second one with 15 unit cells and a Zeeman field $b_x = 0.1$. ($\mu, ~
t_2$ and $b_x$ are all in units of $\ga$).} \label{fig91} \end{figure}

Finally we consider a strip of graphene with a width of 15 unit
cells and a SOC of strength $t_2 = 0.05$, followed immediately by
another strip with the same width of 15 unit cells where there is a
Zeeman field $b_ x = 0.1$ but no SOC; there is pristine graphene on
both sides of the two strips. We consider an incident electron which
first strikes the region with SOC and we study the transmission
after it leaves the region with a Zeeman field. Once again $T_{\ua
\ua}$, $T_{\da \ua}$, $T_{\da \da}$, and $T_{\ua \da}$ will all be
non-zero in general, and the symmetries in Eqs.~\eqref{sym3} will hold.
In Fig.~\ref{fig91}, we show $G_{\ua \ua} /G_0$ and $G_{\da \ua} /G_0$
as functions of $\mu$.

\subsection{Discussion of spin active junctions}
\label{sec:spinactive}

To summarize and compare the results presented in Secs.~\ref{sec:condan}
and \ref{sec:condnum}, we have discussed two kinds of junctions
which are spin active, i.e., they can rotate the direction of spin of
an electron which is incident on the junction. The first example, discussed
in Sec.~\ref{sec:condan}, is a region of graphene which has a combination
of Kane-Mele and Rashba SOC. The Rashba SOC does not conserve the spin;
hence it is not unexpected that it can give rise to a spin active junction of
spin-orbit coupled and pristine graphene. We have used a continuum theory
(valid near the $\vec K$ and $\vec K'$ points) to analytically calculate the
reflection probability from a junction of spin-orbit coupled and pristine
graphene and the transmission probability and differential conductance
(obtained by integrating the transmission over all incident momenta)
through a strip of spin-orbit coupled graphene. To quantify the spin active
nature, we have studied the
amount of spin rotation as a function of the applied voltage and the strength
of the Rashba SOC. In the second example, discussed in Sec.~\ref{sec:condnum},
we studied the effect of a strip of graphene with Kane-Mele SOC, a Zeeman
field (in a direction perpendicular to the SOC so that the two terms
do not commute), and a potential
which is not parity symmetric. Since the calculation cannot be analytically
done for a general non-parity symmetric potential, we have used the
tight-binding model to numerically calculate the transmission probability
and difference conductance across such junctions. Once again we find that
the junction is generally spin active.

Comparing the results for the different junctions, we see that a Zeeman
field which is perpendicular to the Kane-Mele SOC and a Rashba SOC are
most effective in producing spin active junctions. A non-parity symmetric
potential along with a Kane-Mele SOC is relatively less effective.

\section{Discussion}
\label{sec:diss}

In this work, we have studied edge states, effects of impurities, and
spin active junctions in spin-orbit coupled graphene along with the
presence/absence of a Zeeman term which originates from proximity of
the graphene sheet to a suitably chosen ferromagnetic film. The SOC, which
may arise due to proximity of graphene to topological insulator films, has
been taken to be of either Kane-Mele or Rashba form.

Our study concentrates on three properties of such graphene systems.
First, we have shown that a junction between graphene with
the Kane-Mele type of SOC and pristine gapless graphene
(with no SOC) supports robust chiral edge states provided
that the edge separating the two regions is of the zigzag type; no
such states exist for the armchair edges. We have also shown that
these edge states are robust in spite of the presence of gapless
pristine graphene on one side of the junction; this robustness
arises due to the fact that the decay length of these states vanishes
for $t_2 \to 0$. We have pointed out that such behavior is in
complete contrast to the behavior of conventional edge modes where the
decay length diverges in the limit of vanishing gap.

Second, we have studied the change in the LDOS originating from either 
a single or a specific distribution of impurity atoms in spin-orbit coupled
graphene. We have shown that for a single impurity the Fourier
transform of the LDOS displays peaks near the Dirac points with a
finite width; the width of these peaks is a direct measure of the
strength of the induced spin-orbit interaction. We have also shown
that for a specific distribution of impurity atoms (distributed at
the corners of a graphene hexagon), the Fourier transform of LDOS
exhibits an absence of peaks near the Dirac points. Such an absence
can be traced back to the destructive interference of the
contribution to the LDOS from each of the impurity sites and is a
direct signature of the Dirac nature of graphene electrons. Such an
effect has been discussed earlier in the context of LDOS~\cite{bena08}
and STM spectra of single impurity placed at the center of a hexagon in
graphene~\cite{saha10}; however, its manifestation has not been pointed out
for a distribution of impurities to the best of our knowledge.

Although we have only discussed the effects of a single impurity or 
a small number of impurities in this paper, our results can also be used 
to understand what would happen if
there was a finite density of impurities which are far from each other, so 
that the scattering from the different impurities is incoherent. The Fourier 
transform of the change in the LDOS would then be given by the Fourier 
transform of the LDOS for a single impurity multiplied by the density of 
impurities. Hence the Fourier transform of the LDOS of a finite 
density of impurities will share the features of the Fourier transform of 
the LDOS of a single impurity such as the peaks at the Dirac points. The 
Fourier transform of the LDOS for a finite density of impurities can be 
measured by a light scattering experiment.

Third, we have studied junctions of spin-orbit coupled graphene (with both
Kane-Mele and Rashba terms) and pristine graphene. We have shown
that such junctions are generally spin active and that they may be
used to generate electrically controllable spin currents in graphene. We
have demonstrated this in a variety of junctions with analytic computations
using low-energy effective Dirac-like Hamiltonians and with numerical
calculations based on microscopic lattice models. We have also discussed
several ways of enhancing the spin current and pointed out the role
of Zeeman coupling terms and parity-symmetry breaking potential terms
in this context.

The experimental verification of our work would involve preparation
of graphene samples with strong SOC. Since the
intrinsic SOC in graphene is extremely weak, this
needs to be done using a proximate material with strong SOC;
the hybrid samples of topological insulators atop a
graphene sheet which have already been experimentally studied are
ideal for this purpose. The LDOS for impurities in such samples can
be measured using an STM; the Fourier transform of the LDOS can
then be computed~\cite{schouteden}. The prediction of our present work is
that the width of the peak of the Fourier transform of this LDOS
would be a direct measure of the strength of the induced SOC.
To form junctions of spin-orbit coupled graphene with its
pristine counterpart, we need to deposit the topological insulator
over a part of the graphene sample leaving the rest of the sample in
its pristine form. We predict that if such a junction has a zigzag
edge separating the spin-orbit coupled and pristine graphene, there
would be additional chiral edge states whose density of states could
be measured by STM; no such edge states would exist if an armchair
edge separates the two regions. Finally, for spin active junctions
we suggest measurement of the spin current via standard tunneling
conductance measurements where the injection and detection of current
is done with spin-polarized leads of opposite polarities. This will serve
as a direct measure of $G_{\ua \da}$.

To conclude, we have studied edge states, effects of impurities, and spin
active junctions in graphene which has spin-orbit coupling. Our results
points out the presence of robust chiral edge states in graphene junctions
separating spin-orbit coupled graphene from its pristine counterpart with
novel properties of their decay length, shows that the local density of
states originating from impurities in spin-orbit coupled graphene near the
Dirac points can serve as a measure of the strength of the induced
spin-orbit coupling for graphene, and demonstrates that junctions of
spin-orbit coupled and pristine graphene are spin active and may be used
to generate electrically controllable neutral spin currents.
We have proposed realistic experiments which may test our theory.

\acknowledgments
We thank Arindam Ghosh, Kimberly Hsieh and Abhiram Soori for discussions.
D.S. thanks DST, India for support under Grant No. SR/S2/JCB-44/2010.

\end{document}